\PassOptionsToPackage{unicode}{hyperref}
\PassOptionsToPackage{hyphens}{url}
\PassOptionsToPackage{dvipsnames,svgnames,x11names}{xcolor}
\documentclass[
  12pt]{article}

\usepackage{amsmath,amssymb}
\usepackage{iftex}
\ifPDFTeX
  \usepackage[T1]{fontenc}
  \usepackage[utf8]{inputenc}
  \usepackage{textcomp} 
\else 
  \usepackage{unicode-math}
  \defaultfontfeatures{Scale=MatchLowercase}
  \defaultfontfeatures[\rmfamily]{Ligatures=TeX,Scale=1}
\fi
\usepackage{lmodern}
\ifPDFTeX\else
\fi
\IfFileExists{upquote.sty}{\usepackage{upquote}}{}
\IfFileExists{microtype.sty}{
  \usepackage[]{microtype}
  \UseMicrotypeSet[protrusion]{basicmath} 
}{}
\makeatletter
\@ifundefined{KOMAClassName}{
  \IfFileExists{parskip.sty}{%
    \usepackage{parskip}
  }{
    \setlength{\parindent}{0pt}
    \setlength{\parskip}{6pt plus 2pt minus 1pt}}
}{
  \KOMAoptions{parskip=half}}
\makeatother
\usepackage{xcolor}
\makeatletter
\ifx\paragraph\undefined\else
  \let\oldparagraph\paragraph
  \renewcommand{\paragraph}{
    \@ifstar
      \xxxParagraphStar
      \xxxParagraphNoStar
  }
  \newcommand{\xxxParagraphStar}[1]{\oldparagraph*{#1}\mbox{}}
  \newcommand{\xxxParagraphNoStar}[1]{\oldparagraph{#1}\mbox{}}
\fi
\ifx\subparagraph\undefined\else
  \let\oldsubparagraph\subparagraph
  \renewcommand{\subparagraph}{
    \@ifstar
      \xxxSubParagraphStar
      \xxxSubParagraphNoStar
  }
  \newcommand{\xxxSubParagraphStar}[1]{\oldsubparagraph*{#1}\mbox{}}
  \newcommand{\xxxSubParagraphNoStar}[1]{\oldsubparagraph{#1}\mbox{}}
\fi
\makeatother

\usepackage{longtable,booktabs,array}
\usepackage{calc} 
\usepackage{etoolbox}
\makeatletter
\patchcmd\longtable{\par}{\if@noskipsec\mbox{}\fi\par}{}{}
\makeatother
\IfFileExists{footnotehyper.sty}{\usepackage{footnotehyper}}{\usepackage{footnote}}
\makesavenoteenv{longtable}
\usepackage{graphicx}
\makeatletter
\def\maxwidth{\ifdim\Gin@nat@width>\linewidth\linewidth\else\Gin@nat@width\fi}
\def\maxheight{\ifdim\Gin@nat@height>\textheight\textheight\else\Gin@nat@height\fi}
\makeatother

\setkeys{Gin}{width=\maxwidth,height=\maxheight,keepaspectratio}
\makeatletter
\def\fps@figure{htbp}
\makeatother

\makeatletter
\@ifpackageloaded{caption}{}{\usepackage{caption}}
\AtBeginDocument{%
\ifdefined\contentsname
  \renewcommand*\contentsname{Table of contents}
\else
  \newcommand\contentsname{Table of contents}
\fi
\ifdefined\listfigurename
  \renewcommand*\listfigurename{List of Figures}
\else
  \newcommand\listfigurename{List of Figures}
\fi
\ifdefined\listtablename
  \renewcommand*\listtablename{List of Tables}
\else
  \newcommand\listtablename{List of Tables}
\fi
\ifdefined\figurename
  \renewcommand*\figurename{Figure}
\else
  \newcommand\figurename{Figure}
\fi
\ifdefined\tablename
  \renewcommand*\tablename{Table}
\else
  \newcommand\tablename{Table}
\fi
}
\@ifpackageloaded{float}{}{\usepackage{float}}
\floatstyle{ruled}
\@ifundefined{c@chapter}{\newfloat{codelisting}{h}{lop}}{\newfloat{codelisting}{h}{lop}[chapter]}
\floatname{codelisting}{Listing}

\makeatother
\makeatletter
\@ifpackageloaded{caption}{}{\usepackage{caption}}
\@ifpackageloaded{subcaption}{}{\usepackage{subcaption}}
\makeatother

\ifLuaTeX
  \usepackage{selnolig}  
\fi
\usepackage[]{natbib}
\usepackage{bookmark}
\usepackage{rotating}
\IfFileExists{xurl.sty}{\usepackage{xurl}}{} 
\hypersetup{
  pdftitle={Title},
  pdfauthor={Author 1; Author 2},
  pdfkeywords={3 to 6 keywords, that do not appear in the title},
  colorlinks=true,
  linkcolor={blue},
  filecolor={Maroon},
  citecolor={Blue},
  urlcolor={Blue},
  pdfcreator={LaTeX via pandoc}}

  \usepackage{color}
\usepackage{bm}
\usepackage{amsmath}
\usepackage{amssymb}
\usepackage{lscape}
\usepackage{graphicx}
\usepackage{booktabs}
\usepackage{tabularx}
\usepackage{natbib}
\makeatletter
\newtheorem{theorem}{Theorem}\newtheorem{proposition}{Proposition}
\newtheorem{example}{Example}\newtheorem{remark}{Remark}\newtheorem{assum}{Assumption}

  \usepackage{tikz}
\usepackage{algorithm}
\usepackage{algorithmic}
\makeatother

\usepackage{float}
\usepackage{longtable}
\usepackage{natbib}

\newcommand{\anon}{1}

\newcommand{\be}{\mbox{\boldmath $e$}}

\usepackage[bottom]{footmisc}
\begin{document}

\def\spacingset#1{\renewcommand{\baselinestretch}%
{#1}\small\normalsize} \spacingset{1}

\defcitealias{Duan2023}{Duan, Bai and Han}

\defcitealias{Baltagi2021}{Baltagi, Kao and Wang}

\if1\anon
{
  \title{\bf Taxonomy and Estimation of Multiple Breakpoints in High-Dimensional Factor Models}
  \author{Jiangtao Duan\\
    School of Mathematics and Statistics, Xidian University\\
    Jushan Bai\\
    Department of Economics, Columbia University\\
       Xu Han\\
       Department of Economics and Finance, City University of Hong Kong
        }
  \maketitle
} \fi

\if0\anon
{
  \bigskip
  \bigskip
  \bigskip
  \begin{center}
    {\LARGE\bf Title}
\end{center}
  \medskip
} \fi

\bigskip
\begin{abstract}
This paper proposes a quasi-maximum likelihood (QML) estimator for break points in high-dimensional factor models, specifically accounting for multiple
structural breaks. We begin by establishing a necessary and sufficient condition to
categorize two distinct types of breaks in factor loadings: singular changes and rotational changes. The analysis of the nearly singular subsample covariance matrices of the pseudo-factors plays a key role in our approach. It allows us to demonstrate that the QML estimator precisely identifies the true breakpoint with probability tending to one for singular changes. For rotational changes, we demonstrate that the estimator exhibits stochastically bounded estimation errors, implying break fraction consistency.
 Furthermore, we introduce an information criterion to estimate the number
 of breaks, proving that it can detect the true number with probability tending to
 one. Monte Carlo simulations confirm the strong finite sample
 performance of our proposed methods. Finally, we provide an empirical example to
 estimate structural breakpoints in the FRED-MD dataset spanning 1959 to 2024.
\end{abstract}

\noindent%
{\it Keywords:} Quasi-Maximum Likelihood, Singularity, Consistency, Information criterion.
\vfill

\newpage
\spacingset{1.6} 

\vspace{-1em}
\section{Introduction}
\vspace{-1em}

Factor models have become a cornerstone in the analysis of large-scale economic
and financial datasets characterized by both high cross-sectional and time dimensions. The models' versatility has led to widespread applications across economics
and statistics, including the study of policy shocks (\citeauthor{Stock2016}, \citeyear{Stock2016}; \citeauthor{Han2025}, \citeyear{Han2025}), macroeconomic dynamics (\citeauthor{Boivin2006}, \citeyear{Boivin2006}), asset pricing (\citeauthor{Giglio2021}, \citeyear{Giglio2021}; \citeauthor{Giglio2025}, \citeyear{Giglio2025}), portfolio theory (\citeauthor{ross1976}, \citeyear{ross1976}; \citeauthor{chamberlain1983}, \citeyear{chamberlain1983}; \citeauthor{connor1986}, \citeyear{connor1986}, \citeyear{connor1988}; \citeauthor{Ding2021}, \citeyear{Ding2021}),
causal inference (
\citeauthor{Gobillon2016}, \citeyear{Gobillon2016}; \citeauthor{Xu2017}, \citeyear{Xu2017}), and high-dimensional statistical machine learning (\citeauthor{Fan2021}, \citeyear{Fan2021}). However, despite their wide applicability, real-world data are rarely stable over time. Structural breaks are ubiquitous, arising
from policy interventions, financial crises, technological progress, or global events such as the COVID-19 pandemic. Importantly, they often embody the causal impact of interventions and shocks---policies may reshape economic structures, innovations may alter production and financial linkages, and major crises may transform
underlying relationships. Identifying and modeling these breaks is therefore a powerful tool for causal inference, policy evaluation, and understanding how economic
systems evolve in response to major events.

This paper develops a quasi-maximum likelihood (QML) approach for estimating breakpoints in high-dimensional factor models subject to multiple structural
changes. We develop a unified theory that classifies structural breaks in factor loadings into two categories: singular changes and rotational changes. Our asymptotic
results show that the proposed estimator precisely recovers the true breakpoints under singular changes, while for rotational changes the estimation error remains
stochastically bounded. In addition, we propose an information criterion that consistently identifies the true number of breaks with probability approaching one as
the sample size grows.

Previous studies (e.g., \citeauthor{Han2015}, \citeyear{Han2015}) have shown that a single structural break in the factor loading matrix admits an observationally equivalent representation
with time-invariant loadings and an expanded set of pseudo-factors. This expansion induces singularity in the covariance matrices of the pseudo-factors from either the
pre-break or post-break subsample. \citetalias{Duan2023} (\citeyear{Duan2023}, hereafter DBH) exploited this property and demonstrated that such singularity is key to establishing the consistency of the QML estimator in the single-break setting. However, the DBH approach is restricted by its single-break assumption. In comparison, our framework allows for multiple breaks and establishes a necessary and sufficient condition for break taxonomy. This is a critical advancement, as singularity alone no longer suffices to distinguish break types in multi-break settings. Moreover, we develop a novel proof strategy to overcome the technical challenges unique to multi-break setups and propose an information criterion that consistently selects the number of breaks.

Partly due to the simultaneous existence of both singular and rotational breaks, extending the analysis from a single-break to a multi-break setting presents several significant and non-intuitive challenges.
First,
unlike the single-break case, merely checking the singularity of the
pseudo-factor covariance matrix within a given regime is insufficient
to characterize the type of break in a multi-break environment. A
break can be rotational even if neither the preceding nor the
succeeding regime exhibits full column rank, a phenomenon that can
occur due to singular breaks at other time points. Thus, a primary
challenge lies in rigorously defining rotational and singular changes
in factor loadings within a multi-break framework. Second, because
multiple breakpoints are jointly estimated, the estimation errors
of these individual breakpoint estimators become inherently interdependent,
posing substantial technical challenges for asymptotic analysis. Third,
given that the QML estimator exhibits distinct convergence rates for
singular and rotational breaks, consistently determining the true
number of breaks without prior knowledge of their specific types is challenging.

We address these challenges and make the following contributions.
First, we establish a necessary and sufficient condition to precisely
distinguish between singular and rotational structural breaks in factor
loadings within a multi-break framework. Specifically, for each breakpoint,
break types are identified by comparing the numbers of effective factors
in the pre-break and post-break regimes with the number of pseudo-factors in the regime formed by pooling these two regimes. Second,
to tackle the interdependence of estimation errors mentioned previously,
we analyze the asymptotic properties of the $j$-th breakpoint by
``conceptually'' overfitting the remaining true breakpoints. This strategy requires
us to compare the eigenvalues of the sample covariance matrix of the
pseudo-factors in a regime against those from a subsample of that
same regime. Third, leveraging this crucial result, we prove that
the QML estimator identifies the true breakpoint with probability
tending to one in large samples for singular breaks, and achieves
$O_{p}(1)$ estimation errors for rotational breaks. Finally, we analyze
how underfitting and overfitting the number of breaks affects the
value of our QML objective function under both types of breaks. This
analysis enables us to design an effective penalty function for our
information criterion, ensuring its consistent selection of the true
number of breaks without requiring prior knowledge of break types.

Moreover, we perform a series of Monte Carlo simulations to
assess the finite sample properties of our estimator. The results
demonstrate that our method can accurately estimate the break dates,
even in the presence of small sample sizes, and that our information
criterion consistently identifies the correct number of breaks. We
further apply our approach to the FRED-MD dataset (\citeauthor{McCracken2016}, \citeyear{McCracken2016}), detecting five breakpoints over the period from January 1959
to July 2024.

{\bf Links to the literature.}
A growing body of literature investigates tests for structural breaks
in high-dimensional factor models (e.g., \citeauthor{Chen2014}, \citeyear{Chen2014}; \citeauthor{Han2015}, \citeyear{Han2015}; \citeauthor{Yamamoto2015}, \citeyear{Yamamoto2015}; \citeauthor{Bai2024}, \citeyear{Bai2024};  \citeauthor{fan2024tests}, \citeyear{fan2024tests}; \citeauthor{Fan2024}, \citeyear{Fan2024}; \citeauthor{Peng2025}, \citeyear{Peng2025}). Our study differs by focusing on the estimation
of breakpoints rather than testing for their presence. Moreover, traditional
literature on structural break estimation primarily addresses time
series with small cross-sectional dimensions (e.g., \citeauthor{Bai1998}, \citeyear{Bai1998}; \citeauthor{Bai2000}, \citeyear{Bai2000}; \citeauthor{Qu2007}, \citeyear{Qu2007}; \citeauthor{Chan2014}, \citeyear{Chan2014}, \citeyear{Chan2021}). In contrast,
our factor model framework operates in a setting where both the cross-sectional
($N$) and time ($T$) dimensions are large.

Within this large $N$, large $T$ context, prior work has explored
structural breaks in various settings. \cite{Qian2016}, for instance,
applied group fused Lasso to panel models with structural breaks,
and \cite{Li2016} extended this to panel data with interactive
fixed effects. While these studies considered large panels, their focus
differed from ours, as they concentrated on breakpoints in regression
coefficients rather than the high-dimensional factor loading matrix.
For high-dimensional time series, \cite{Bai2010} developed a least squares
(LS) estimator for breakpoints in the mean and variance, and \cite{Li2024} introduced a two-way MOSUM procedure with asymptotic properties
for change-point estimators in trend functions. Our framework distinguishes
itself from these by specifically addressing breakpoints in the loading
coefficients of unobserved factors.

Regarding breakpoint estimation in factor models, earlier literature
established the break fraction consistency, specifically
that $\hat{\tau}-\tau_{0}=o_{p}(1)$, where $\hat{\tau}$ is the estimated
breakpoint $\hat{k}$ divided by sample size $T$, and $\tau_{0}$
is the true fraction (e.g., \citeauthor{Chen2015}, \citeyear{Chen2015}; \citeauthor{Cheng2016}, \citeyear{Cheng2016}). More
recent studies have provided theoretical results on the detailed rate
of the error bound for the estimated breakpoint. For example, \cite{Barigozzi2018} proposed a wavelet-based method to estimate both the
number and locations of structural breaks in common and idiosyncratic
components. \cite{Barigozzi2024} further developed a MOSUM procedure
for detecting multiple change points in high-dimensional factor models.
Both studies showed that the estimation error of the breakpoint diverges
slowly with sample size, implying a consistently estimated break fraction.
Similarly, \citeauthor{Baltagi2017} (\citeyear{Baltagi2017}, \citeyear{Baltagi2021}) developed an estimator that
compares subsample second moments of factors estimated via full-sample
principal components. Their breakpoint estimator exhibits a stochastically
bounded estimation error, which also implies a consistently estimated
break fraction.

In contrast to these approaches, our QML estimator achieves not only
fraction-consistency but also point-consistency (sometimes referred to as super-consistency in the literature), meaning the estimated
breakpoints equal the true ones with probability approaching one in
large samples. To our knowledge, only a few studies have established
point-consistency for breakpoint estimators of the loading matrix
in factor models. \cite{Bai2020} demonstrated that the LS estimator
is point-consistent under both large and small break settings. However,
their framework is limited to a single break, whereas our methodology
accommodates multiple breaks. \cite{Ma2018} proposed a fused Lasso
estimator that is consistent for multiple breakpoints in factor models,
under the assumption of a constant number of factors. Our work considers
a more general and empirically relevant setting by allowing the number
of effective factors to vary over time. Additionally, \cite{Ma2023} adopted a group fused Lasso approach to estimate changes in regression
coefficients of one factor on the remaining factors. Their fused Lasso
estimator exhibits stochastically bounded errors, which are then refined
by LS to achieve point-consistency. Nonetheless, their method does
not accommodate certain types of rotational changes (e.g., when the
post-break loading matrix is rescaled by a constant), a feature that our
Type 2 break in Section 3 does allow.

{\bf Plan of the Paper.} The rest of this paper is organized as follows. Section 2
introduces the factor model with multiple breaks on the factor loading
matrix and describes the QML breakpoint estimator. Section 3 categorizes the two types of breakpoints. Section 4 outlines the assumptions for this model and presents the asymptotic
theory. Section 5 investigates the finite-sample properties of the
QML estimators through simulations. Section 6 provides an empirical
study. Section 7 concludes.

{\bf Notation.} The following notation is used throughout the paper.
Let $\rho_{i}(\mathbb{B})$ denote the $i$-th eigenvalue of an $n\times n$
symmetric matrix $\mathbb{B}$, and $\rho_{1}(\mathbb{B})\geq\rho_{2}(\mathbb{B})\geq\cdots\geq\rho_{n}(\mathbb{B})$.
For an $m\times n$ real matrix $\mathbb{A}$, we denote its Frobenius
norm as $\|\mathbb{A}\|=[tr(\mathbb{A}\mathbb{A}^{'})]^{1/2}$, and
its adjoint matrix as $\mathbb{A}^{\#}$ when $m=n$. Let $\mathrm{Proj}(\mathbb{A}|\mathbb{Z})$
denote the projection of columns of matrix $\mathbb{A}$ onto the columns of
matrix $\mathbb{Z}$. For $a,\ b>0$, $a\gtrsim b$
indicates that there exists a constant $c>0$ such that $a\geq cb$.
We write $a\asymp b$ to mean that there exist constants $c_{1},c_{2}>0$
such that $c_{1}b\leq a\leq c_{2}b$. For a real number $x$, $[x]$ represents
the integer part of $x$.

\begin{samepage}
\vspace{-0.5\baselineskip}

\vspace{-1em}
\section{Model and estimator}
\vspace{-1em}

\vspace{-1em}
\subsection{Model}
\vspace{-1em}

\end{samepage}

We consider the following factor model with $m_{0}$ common breaks at $k_{1}^{0}(T)<k_{2}^{0}(T)<\cdots<k_{m_{0}}^{0}(T)$ in the factor loadings for $i=1,\cdots,N$:
\begin{eqnarray}
x_{it}=\begin{cases}
\lambda_{i,1}'f_{t}+e_{it} & for~~t=1,2,\cdots,k_{1}^{0}(T)\\
\cdots\\
\lambda_{i,m_{0}+1}'f_{t}+e_{it} & for~~t=k_{m_{0}}^{0}(T)+1,\cdots,T,\end{cases}\label{model_1}\end{eqnarray}
where $f_t$ is an $r$-dimensional vector of unobserved common factors; $r$ is the number of pseudo-factors in the entire sample; $k_j^0(T),j=1,\cdots,m_0$ are unknown break dates and $\min_{\{j=0,1,\cdots,m_0\}}k_{j+1}^0(T)-k_{j}^0(T)\geq \eta T$ for a constant $0<\eta<1$, with $k_{0}^0(T)=0$ and $k_{m_0+1}^0(T)=T$; $\lambda_{i,j}$
are $r\times1$ factor loading vectors for unit $i$ in the $j$th regime $t\in[k_{j-1}^0(T)+1,\,k_j^0(T)]$ for $j=1,\cdots,m_0+1$; and $e_{it}$ is the error term allowed to have serial and cross-sectional dependence as well as heteroskedasticity, and both $N$ and $T$ tend to infinity.
$\tau_j^0=k_j^0(T)/T\in (0,1),j=1,\cdots,m_0$ are break fractions and are fixed constants. For notational simplicity, hereinafter, we suppress the dependence of $k_j^0$ on $T$. Note that the
dimension of $f_{t}$ is the same as that of the pseudo-factors (defined below).

In vector form, model (\ref{model_1}) can be expressed as \begin{eqnarray}
x_{t}=\begin{cases}
\Lambda_{1}f_{t}+e_{t} & for~~t=1,2,\cdots,k_{1}^{0}\\
\cdots\\
\Lambda_{m_{0}+1}f_{t}+e_{t} & for~~t=k_{m_{0}}^{0}+1,\cdots,T,\end{cases}\label{model_vector}\end{eqnarray}
 where $x_{t}=[x_{1,t},\cdots,x_{N,t}]^{'}$, $e_{t}=[e_{1,t},\cdots,e_{N,t}]^{'}$,
$\Lambda_{j}=[\lambda_{1,j},\cdots,\lambda_{N,j}]^{'},j=1,\cdots,m_{0}+1$.

For any $j=1,\cdots,m_0+1$, we define
$$X_{j}=[x_{k_{j-1}^0+1},\cdots,x_{k_j^0}]^{'},~~F_{j}=[f_{k_{j-1}^0+1},\cdots,f_{k_j^0}]^{'},~~\be_{j}=[e_{k_{j-1}^0+1},\cdots,e_{k_j^0}]^{'}.$$
We rewrite (\ref{model_vector}) using the following matrix representation:
\begin{eqnarray}
X=\left[\begin{array}{cccccccccc}
X_{1}\\
\cdots\\
X_{m_{0}+1}\end{array}\right] & = & \left[\begin{matrix}F_{1}\Lambda_{1}^{'}\\
\cdots\\
F_{m_{0}+1}\Lambda_{m_{0}+1}^{'}\end{matrix}\right]+\left[\begin{array}{cccccccccc}
\be_{1}\\
\cdots\\
\be_{m_{0}+1}\end{array}\right]=\left[\begin{matrix}F_{1}(\Lambda B_{1})^{'}\\
\cdots\\
F_{m_{0}+1}(\Lambda B_{m_{0}+1})^{'}\end{matrix}\right]+\left[\begin{array}{cccccccccc}
\be_{1}\\
\cdots\\
\be_{m_{0}+1}\end{array}\right],\nonumber \\
 & = & \underbrace{\left[\begin{matrix}F_{1}B_{1}^{'}\\
\cdots\\
F_{m_{0}+1}B_{m_{0}+1}^{'}\end{matrix}\right]}_{G}\Lambda^{'}+\left[\begin{array}{cccccccccc}
\be_{1}\\
\cdots\\
\be_{m_{0}+1}\end{array}\right],\nonumber \\
 & = & G\Lambda^{'}+E,\label{Baltagi}\end{eqnarray}
 where $F_j,j=1,\cdots,m_{0}+1$ has dimensions $(k_{j}^{0}-k_{j-1}^{0})\times r$ with full column rank for $j=1,\cdots,m_0+1$
and $\Lambda$ is an $N\times r$ matrix with full column rank. The
loadings are modeled as $\Lambda_{j}=\Lambda B_{j}$, $j=1,\cdots,m_{0}+1$,
where $B_{j}$ is an $r\times r$ matrix.
Each $\Lambda_{j}$ has dimension $N\times r$. 
In this model, $$r_{j}=rank(B_{j})\leq r$$
denote the numbers of \emph{effective
factors} in regime $j$, i.e., the rank of the common component in regime $[k_{j-1}^0+1,k_j^0]$.
In (\ref{Baltagi}), we define
$$G=[g_1,\cdots,g_T]'$$
as the \emph{pseudo-factors} because the final expression in (\ref{Baltagi}) provides an observationally equivalent representation, maintaining the structure of the loadings matrix \( \Lambda \) unchanged. More precisely, when the break is omitted during the estimation, the factors derived from a full-sample PCA correspond to the pseudo-factors \( G \) presented in (\ref{Baltagi}). It is well established that the presence of breaks can expand the factor space; consequently, for \( j = 1, \dots, m_{0}+1 \), we have \( r_j \leq r \), where \( \text{rank}(G) =\text{rank}([B_1, B_2, \dots, B_{m_0 + 1}])= r \).
This representation allows for changes in factor loadings and the number of factors. Concrete constructions of $\Lambda$, $B_j$, $r$, and $r_j$ under different breakpoint types are given in Examples 1--3 of Section 3.
We differentiate between effective factors and pseudo-factors. Effective factors are the factors required to capture common variation within a single, stable regime. In contrast, pseudo-factors are used to describe the factors (whose dimension may exceed that of effective factors due to breaks) within either the full sample (i.e., $G$) or a subsample consisting of consecutive regimes.

\subsection{Estimator}
\vspace{-1em}

In what follows, we utilize the QML method to estimate
the break dates for Model (\ref{Baltagi}), assuming the number of
breaks is known. An information criterion for determining the number
of breaks is also proposed in Section 4. 

Let $\hat{G}=(\hat{g}_{1},...,\hat{g}_{T})^{\prime}$
denote the full-sample PCA estimator for $G$ subject to the normalization condition that
$G^{'}G/T=I_r$ and $\Lambda^{'}\Lambda$ being diagonal. The number of pseudo-factors can be estimated following \cite{Bai2002}. \footnote{Other methods for determining the number of factors include the eigenvalue ratio and growth ratio (\citeauthor{Ahn2013},  \citeyear{Ahn2013}), a similar method using adjacent eigenvalue ratios proposed by \cite{LamYao2012}, the empirical distribution estimator (\citeauthor{Onatski2010}, \citeyear{Onatski2010}), and the bridge estimator (\citeauthor{Caner2014}, \citeyear{Caner2014}).}
For a given set of splitting points $(k_1,\cdots,k_{m_0})$, the QML objective function
is expressed as follows: \begin{eqnarray*}
U_{NT}(k_{1},\cdots,k_{m_0}) & = & \sum\limits _{\ell=1}^{m_0+1}(k_{\ell}-k_{\ell-1})\log|\hat{\Sigma}(k_{\ell-1},k_{\ell})|,\end{eqnarray*}
 where \[
\hat{\Sigma}(k_{\ell-1},k_{\ell})=\frac{1}{k_{\ell}-k_{\ell-1}}\sum\limits _{t=k_{\ell-1}+1}^{k_{\ell}}\hat{g}_{t}\hat{g}_{t}^{'},\]
with $k_{0}=0$ and $k_{m_0+1}=T$.
 The QML estimator of the breakpoints for model (\ref{Baltagi}) is defined as follows:
\begin{eqnarray}
 &  & (\hat{k}_{1},\cdots,\hat{k}_{m_0})=\arg\min_{(k_1,\cdots,k_{m_0})\in K_\eta }U_{NT}(k_{1},\cdots,k_{m_0}),\label{obj1_fun}\end{eqnarray}
where $K_\eta=\{(k_1,\cdots,k_{m_0}): 0=k_0<k_1<\cdots<k_{m_0}<k_{m_0+1}=T, k_j-k_{j-1} \geq T\eta,j=1,\cdots,m_0+1\}$ and $\eta \in (0,1)$ is a small positive constant.

\begin{remark}[Dynamic programming implementation for the QML estimator]\label{rem:DP_QML}
We explain how to compute the QML estimator in (4) using a
dynamic programming (DP) algorithm in the spirit of Bai and Perron (1998, 2003).
For any integers $0\le s<e\le T$, define
\[
\hat{\Sigma}(s,e)
=
\frac{1}{e-s}\sum_{t=s+1}^{e}\hat g_t\hat g_t',
\qquad
\mathcal Q(s,e)
=
(e-s)\log\big|\hat{\Sigma}(s,e)\big|.
\]
Then for any partition $0=k_0<k_1<\cdots<k_m<k_{m+1}=T$,
the QML objective satisfies the additivity property
\[
U_{NT}(k_1,\ldots,k_m)
=
\sum_{\ell=1}^{m+1}\mathcal Q(k_{\ell-1},k_\ell).
\]
Let $h=\lceil T\eta\rceil$ be the minimum segment length implied by the
constraint set $K_\eta$. Define $D(m,e)$ as the minimum value of
$\sum \mathcal Q(\cdot,\cdot)$ over all partitions of the first e observations $\{1,\ldots,e\}$
with exactly $m$ breaks such that each segment has
length at least $h$. The DP recursion is
\[
D(0,e)=\mathcal Q(0,e),\qquad e\ge h,
\]
and for $m\ge 1$ and $e\ge (m+1)h$,
\[
D(m,e)=\min_{j\in\{mh,\ldots,e-h\}}\Big\{D(m-1,j)+\mathcal Q(j,e)\Big\}.
\]

For a given $m$, evaluating $D(m, e)$ for all $e \ge  (m+1)h$ requires $O(T^2)$ operations by varying the endpoint $e$ and the candidate breakpoint $j \le e-h$. 
The final breakpoint sequence is recovered via backtracking using the stored argmin indices: 
let $\hat{k}_{m_0+1}=T$, and for $\ell=m_0,\ldots,1$ set $\hat{k}_\ell$ equal to the minimizer of $D(\ell-1,j)+\mathcal{Q}(j,\hat{k}_{\ell+1})$ over $j\in\{\ell h,\ldots,\hat{k}_{\ell+1}-h\}$.
 \end{remark}

\begin{algorithm}[H]
\caption{DP computation of $(\hat k_1,\ldots,\hat k_{m_0})$ for a fixed $m_0$}\label{alg:DP_QML}
\begin{algorithmic}[1]
\STATE \textbf{Input:} $\{\hat g_t\}_{t=1}^T$, $\eta\in(0,1)$, $h=\lceil T\eta\rceil$, a fixed number of breaks $m_0$.
\STATE \textbf{Precompute prefix sums:} $S_0=0_{r\times r}$ and $S_t=\sum_{u=1}^t \hat g_u\hat g_u'$ for $t=1,\ldots,T$.
\STATE \textbf{Segment costs:} For all $0\le s<e\le T$ with $e-s\ge h$, set
$\hat\Sigma(s,e)=(S_e-S_s)/(e-s)$ and $\mathcal Q(s,e)=(e-s)\log|\hat\Sigma(s,e)|$.

\vspace{0.3em}
\STATE \textbf{DP initialization:} For $e=h,\ldots,T$, set $D(0,e)=\mathcal Q(0,e)$.
\FOR{$m=1,\ldots,m_0$}
  \FOR{$e=(m+1)h,\ldots,T$}
    \STATE $D(m,e)=\min\limits_{j=mh,\ldots,e-h}\{D(m-1,j)+\mathcal Q(j,e)\}$.
\STATE Store an argmin value:  $A(m,e)=\arg\min\limits_{j=mh,\ldots,e-h}\{D(m-1,j)+\mathcal Q(j,e)\}$.  
  \ENDFOR
  
\ENDFOR

\vspace{0.3em}
\STATE \textbf{Backtracking:} Set $\hat k_{m_0+1}=T$ and $\hat k_{\ell}=A(\ell,\hat k_{\ell+1})$ for $\ell=m_0,\ldots,1$.
\STATE \textbf{Output:} $(\hat k_1,\ldots,\hat k_{m_0})$.
\end{algorithmic}
\end{algorithm}

The estimator can be viewed as a two-step procedure: first estimate $\hat G$ by full-sample PCA, and then locate the breakpoints by minimizing the QML objective function. 
Unlike classical methods, our log-determinant objective exploits the near-singularity of $\hat{\Sigma}(k_{\ell-1}, k_{\ell})$ when the segment boundaries approach true Type 1 breakpoints.
This sensitivity to the rank-deficient structure enables the faster convergence rates established in Theorem 2, outperforming existing methods such as \cite{Barigozzi2018, Barigozzi2024}, \cite{Baltagi2017, Baltagi2021}.

Once the breakpoints are estimated, one can apply split-sample PCA to consistently estimate the loadings and factors for each regime. Using data within the interval $[\hat{k}_{j-1}+1, \hat{k}_j]$, one can recover the regime-specific $r_j$-dimensional factors $\hat{f}_t$ and loadings $\hat{\Lambda}_j$,  subject to a local $r_j \times r_j$ rotation $\hat{H}_j$.

In the next section, we classify breakpoints into two types based on the rank properties of the loading matrices across regimes.

\vspace{-1em}
\section{Types of breakpoints}
\vspace{-1em}
The representation in (\ref{Baltagi}) is convenient for theoretical analysis because one can control break types by setting $B_j$ for $j=1,\cdots,m_0+1$.
To further illustrate the generality of model (\ref{Baltagi}), we present two types of breakpoint structures within the combined time interval \([k_{j-1}^0+1, k_{j+1}^0]\) (combined regime) by varying the ranks of \( B_j \) and \( B_{j+1} \). We define
the number of pseudo-factors in the combined regime $[k_{j-1}^{0}+1,k_{j+1}^{0}]$ as
\footnote{This is because\begin{align*}
r_{j,j+1} & =\text{rank}([B_{j}F_{j}',B_{j+1}F_{j+1}']')=\text{rank}([B_{j}F_{j}',B_{j+1}F_{j+1}'][B_{j}F_{j}',B_{j+1}F_{j+1}']')\\
 & =\text{rank}([B_{j},B_{j+1}]diag(F_{j}'F_{j},\ F_{j+1}'F_{j+1})[B_{j},B_{j+1}]')=\text{rank}([B_{j},B_{j+1}]).\end{align*}
}\[
r_{j,j+1}=\text{rank}([B_{j}F_{j}',B_{j+1}F_{j+1}']')=\text{rank}([B_{j},B_{j+1}]).\]
Next, we show that there are two types of breakpoints.

\textbf{Type 1}. $r_{j,j+1}>\min\{r_{j,}r_{j+1}\}$. In this case, the number of pseudo-factors in the
combined regime $[k_{j-1}^{0}+1,k_{j+1}^{0}]$ exceeds the number
of effective factors in at least one of the two regimes: $[k_{j-1}^{0}+1,k_{j}^{0}]$
and $[k_{j}^{0}+1,k_{j+1}^{0}]$. We refer to this as {\bf a
singular change}.

If $r_{j,j+1}>\max\{r_{j},r_{j+1}\}$, the dimension
of the factor space over the combined regime $[k_{j-1}^{0}+1,k_{j+1}^{0}]$
increases due to the structural break. This occurs when the factor
loading matrices from regime $[k_{j-1}^{0}+1,k_{j}^{0}]$ and regime
$[k_{j}^{0}+1,k_{j+1}^{0}]$ jointly span a larger space than either
individual matrix. This space augmentation is analogous to a regression
where a break in coefficients increases the number of effective regressors;
specifically, the interaction of a break dummy with original regressors
leads to an expanded regressor set. 

\begin{example} 
Consider a three-regime design where each regime $j$ has $r_j = 3$ effective factors and a loading matrix $\breve{\Lambda}_j$. Assume that the column spaces of  $\{\breve{\Lambda}_j\}_{j=1}^3$ are linearly independent. To see how our framework accommodates this setting,  we set $r=9$ and $\Lambda = [\breve{\Lambda}_1, \breve{\Lambda}_2, \breve{\Lambda}_3]$ and define $B_j$ as a $9 \times 9$ selection matrix with $I_3$ as the $j$-th diagonal block and zeros elsewhere. Thus, $\{\Lambda B_j\}_{j=1}^3$ collect the first, middle, and last three columns of $\Lambda$. Under this setup, any two adjacent regimes jointly span a six-dimensional factor space, implying $r_{j,j+1} = 6 > \max\{r_j, r_{j+1}\} = 3$ for $j=1,2$, so both breakpoints are Type 1 (singular) changes. 
\end{example}

\begin{example} 
Consider a scenario with $r=3$ pseudo-factors and loading matrix $\Lambda_{N \times 3}$. Regime $j$ has three effective factors, and in regime $j+1$ one factor becomes inactive after $k_j^0$, reducing the effective rank to $r_{j+1}=2$. 
Our framework in (3) accommodates this case by setting $B_j = I_3$ and $B_{j+1} = \text{diag}(1, 1, 0)$. Since the column space of $\Lambda B_{j+1}$ is a proper subspace of that of $\Lambda B_j$, we have $r_{j,j+1} = 3 = \max\{r_j, r_{j+1}\} > \min\{r_j, r_{j+1}\} = 2$, which identifies the breakpoint as a Type 1 (singular) change.
\end{example}

\textbf{Type 2}. $r_{j,j+1}=\min\{r_j,r_{j+1}\}$. In this case, the number of pseudo-factors in the combined regime \([k_{j-1}^0+1, k_{j+1}^0]\) is the same as the number of effective factors in each regime, and we refer to this as {\bf a rotational change}. Specifically, there exists a nonsingular $r\times r$ matrix $\bf R$ such that
\begin{equation}
B_{j}{\bf R}=B_{j+1},\label{eq:Bj R}\end{equation}
then
\begin{eqnarray}\label{rotational_defi}
\left[\begin{array}{cccccccccc}
F_jB_j'\\
F_{j+1}B_{j+1}'\end{array}\right]=\left[\begin{array}{cccccccccc}
F_jB_j'\\
F_{j+1}{\bf R}'B_{j}'\end{array}\right]=\left[\begin{array}{cccccccccc}
F_j\\
F_{j+1}{\bf R}'\end{array}\right]B_j',\end{eqnarray}
which satisfies $r_{j,j+1}=r_j=r_{j+1}=rank(B_j)$. When there is one break ($m_0=1$) and $B_1$ is nonsingular, it is analogous to the rotational change for the single break setup. 

\begin{example} 
Consider $r=3$ pseudo-factors where the loading matrix in regime $j+1$ is exactly twice that of regime $j$. Our framework accommodates this by setting $B_j = I_3$ and $B_{j+1} = 2I_3$, such that $\Lambda B_{j+1} = 2\Lambda B_j$. Because this nonsingular scaling preserves the column space, both regimes span the same three-dimensional factor space. Hence, $r_{j,j+1} = r_j = r_{j+1} = 3$, identifying the breakpoint as Type 2 (rotational). More generally, the same conclusion holds if $B_{j+1} = B_j\textbf{R}$ for a nonsingular $3 \times 3$ matrix $\textbf{R}$, where the break can equivalently be interpreted as a change in the factor covariance matrix with an unchanged loading space.
\end{example}

\begin{remark}
A breakpoint is classified as a Type 1 (singular)
change if $\min\{r_{j},r_{j+1}\}<r_{j,j+1}$.
This inequality means that the common factor loading spaces of the two adjacent regimes are distinct. These spaces could partially overlap, one might be nested within the other, or they could be entirely separate. Regardless, the dimension of the union of these spaces ($r_{j,j+1}$) will always be strictly greater than the minimum of the individual regime ranks ($\min\{r_{j},r_{j+1}\}$). This leads to singularity in either $B_j$ or $B_{j+1}$, hence the designation singular change.
Such a change typically suggests that the factor loadings have shifted in a non-proportional manner across the breakpoint.
Conversely, we define a breakpoint
as a Type 2 (rotational) change if $r_{j}=r_{j+1}=r_{j,j+1}$. In
this scenario, the loadings change in a way such that the pre- and post-break loading matrices span the same space. This ensures that the factor dimension remains constant across the breakpoint.
In the specific case of a single breakpoint, this classification simplifies
to the rule established by DBH: if either $B_{1}$ or $B_{2}$ is
singular, the breakpoint represents a singular (Type 1) change; if
both are full rank, it is a purely rotational (Type 2) change.\end{remark}

These two types of breaks have distinct economic interpretations. Rotational (Type 2) breaks can be interpreted as observationally equivalent to a change in the second moment of the existing factors, while leaving the loading space unchanged. In contrast, singular (Type 1) breaks represent ``heterogeneous'' shifts where the column space of the loading matrix changes. This corresponds to more profound structural transformations, including the emergence of new factors or the disappearance of existing ones.

\begin{remark}\label{rotational_singular_rmk}
Notably, the analysis of multiple breakpoints
is considerably more complicated than \emph{for
a single breakpoint.} In a multi-breakpoint
setting, $r_{j,j+1}$ can be less than the
total number of pseudo-factors $r$. Specifically,
it is possible that both $B_{j}$ and $B_{j+1}$ are singular, yet
the $j$-th breakpoint is rotational. Consequently, relying solely
on the singularity of $B_{j}$ (like the
DBH rule) is insufficient for determining
the breakpoint type. To understand this
phenomenon, consider Equations (\ref{eq:Bj R})
and (\ref{rotational_defi}): when $B_{j}$
is singular (and consequently $B_{j+1}$ is singular),
the pseudo-factors in regimes $j$ and $j+1$ (i.e., $F_{j}B_{j}'$
and $F_{j+1}B_{j+1}'$) are not of full
column rank. However, if we focus solely
on the data in the combined interval $[k_{j-1}^{0}+1,k_{j+1}^{0}]$
and fit a factor model for observations
within this interval, the pseudo-factors in this model are of full
column rank $r_{j,j+1}$. In
contrast, when considering the full sample for  $t=1,...,T$,
the pseudo-factors in the interval $[k_{j-1}^{0}+1,k_{j+1}^{0}]$
become rank-deficient.
Therefore, this singularity is not directly attributable to the $j$-th
breakpoint itself but rather results from the expansion of the factor
loading space caused by other breakpoints, which renders
the pseudo-factors in $[k_{j-1}^{0}+1,k_{j+1}^{0}]$
rank-deficient. This highlights that for
rotational changes, the singularity of
$B_{j}$ and $B_{j+1}$ is not introduced by the $j$-th breakpoint
but rather by the cumulative effect of structural changes at multiple
locations. Therefore,
relying solely on the singularity of $B_{j}$
and $B_{j+1}$ is
insufficient to distinguish between Type 1 and Type 2 changes at the
$j$-th breakpoint, as both types may be associated with singular
$B_{j}$ and $B_{j+1}$. Situations involving rank
deficiency for regimes before and after
a rotational change occur only when multiple breakpoints are present;
such scenarios do not arise in single-breakpoint cases, highlighting
the complexity introduced by the multi-breakpoint setting.
\end{remark}

The preceding discussion naturally raises the question
of how to distinguish between the two types of changes in multi-breakpoint
settings, particularly when the singularity of $B_{j}$ and $B_{j+1}$
(the DBH rule in a single-break setting) is no longer sufficient for
distinction. The answer is to use the definition $r_{j}=r_{j+1}=r_{j,j+1}$
to determine whether the $j$-th breakpoint is rotational. Evidently, if
Equation (\ref{eq:Bj R}) holds, then $r_{j}=r_{j+1}=r_{j,j+1}$ must
also hold, implying that the $j$-th breakpoint is rotational. Conversely,
another natural question arises: does $r_{j}=r_{j+1}=r_{j,j+1}$ always
imply the existence of a nonsingular $R$ matrix satisfying Equation
(\ref{eq:Bj R})? This is affirmed by Proposition \ref{definite_rotational_change},
which demonstrates that Equation (\ref{eq:Bj R}) is a necessary and
sufficient condition for defining a Type 2 break. This proposition
also explains why Type 2 is named a {}``rotational change.''

\begin{proposition}
\label{definite_rotational_change}
There exists a nonsingular $\bf R$ such that $B_j{\bf R}=B_{j+1}$ $\Leftrightarrow$ $r_{j,j+1}=\min(r_j,r_{j+1})$.

\end{proposition}

Proposition \ref{definite_rotational_change} clarifies that
the condition $B_{j}{\bf R}=B_{j+1}$ is equivalent to the definition
of a Type 2 breakpoint (i.e., $r_{j}=r_{j+1}=r_{j,j+1}$). In practice,
researchers do not need to directly verify the existence of ${\bf R}$
such that $B_{j}{\bf R}=B_{j+1}$ is satisfied. Instead, based on
the conclusion of Proposition \ref{definite_rotational_change}, the
breakpoint type can be readily determined by simply comparing the
magnitudes of $r_{j}$, $r_{j+1}$, and $r_{j,j+1}$. If these quantities
are equal, it indicates a Type 2 change; otherwise, it is a Type 1
change.

\begin{remark} Proposition \ref{definite_rotational_change}
can equivalently be written as follows: There does not exist a nonsingular
${\bf R}$ such that $B_{j}{\bf R}=B_{j+1}$ $\Leftrightarrow$ $r_{j,j+1}>\min(r_{j},r_{j+1})$,
which corresponds to a type 1 change. To
further illustrate this, consider two distinct cases for a setting
involving two breakpoints. Case 1 involves three regimes characterized
by $B_{1}$, $B_{2}$, and $B_{3}$; Case 2 involves $B_{1}$, $\tilde{B}_{2}$,
and $B_{3}$. These matrices are defined as follows: \begin{eqnarray*}
B_{1}=\left[\begin{array}{ccc}
1,\ 0,\ 0\\
0,\ 0,\ 0\\
0,\ 0,\ 0\end{array}\right],\,\, B_{2}=\left[\begin{array}{ccc}
1,\ 1,\ 0\\
0,\ 0,\ 0\\
0,\ 0,\ 0\end{array}\right],\,\,\tilde{B}_{2}=\left[\begin{array}{ccc}
1,\ 0,\ 0\\
1,\ 0,\ 0\\
0,\ 0,\ 0\end{array}\right],\,\, B_{3}=I_{3},\end{eqnarray*}
 so rank$(B_{1})$=rank$(B_{2})$=rank$(\tilde{B}_{2})=1$. But
rank$(\left[B_{1},B_{2}\right])=1$ and rank$(\left[B_{1},\tilde{B}_{2}\right])=2>rank(B_{1})$.
For $B_{1}$, we can find a nonsingular $r\times r$ matrix ${\bf R}$
such that \begin{eqnarray*}
B_{1}\times{\bf R}=\left[\begin{array}{ccc}
1,\ 0,\ 0\\
0,\ 0,\ 0\\
0,\ 0,\ 0\end{array}\right]\times\left[\begin{array}{ccc}
1,\ 1,\ 0\\
0,\ 1,\ 0\\
0,\ 0,\ 1\end{array}\right]=\left[\begin{array}{ccc}
1,\ 1,\ 0\\
0,\ 0,\ 0\\
0,\ 0,\ 0\end{array}\right]=B_{2}.\end{eqnarray*}
 However, such nonsingular ${\bf R}$ does not exist for $\tilde{B}_{2}$
because the second row of the product $B_{1}{\bf R}$ is always zeros,
which is not equal to that of $\tilde{B}_{2}$, i.e., for any ${\bf R}$,
\begin{eqnarray*}
B_{1}\times{\bf R}\neq\left[\begin{array}{ccc}
1,\ 0,\ 0\\
1,\ 0,\ 0\\
0,\ 0,\ 0\end{array}\right]=\tilde{B}_{2}.\end{eqnarray*}
Thus, in this example, for Case 1, where
$B_{2}$ is used, since $r_{1,2}=r_{1}=r_{2}=1$, the first breakpoint
is classified as Type 2, i.e., a rotational change. In contrast, for
Case 2, where $\tilde{B}_{2}$ is used, since $r_{1,2}=2>\min\{r_{1},r_{2}\}=1$,
this first breakpoint is classified as Type 1, i.e., a singular change.
Furthermore, in both Case 1 and Case 2, the second breakpoint is a
singular change because $r_{2,3}=3>\min\{r_{2},r_{3}\}$.
\par \label{definite_rotational_change_remark} \end{remark}

For Type 1 breakpoints, we will show that the QML estimators converge to the true breakpoints with probability approaching one. In contrast, for Type 2 breakpoints, the difference between QML estimators and the true change points is stochastically bounded. The specific asymptotic theory is detailed in Section 4.
\vspace{-1em}
\section{Asymptotics of the estimators}
\vspace{-1em}
\vspace{-1em}
\subsection{Assumptions}
\vspace{-1em}
The assumptions are as follows:
\begin{assum}\label{factors}

(i) $E\left\|f_t \right\|^4<M<\infty$, $E(f_tf_t^{'})=\Sigma_F$, where $M$ is a positive constant, $\Sigma_F$ is positive definite, and
$\frac{1}{k_j^0-k_{j-1}^0}\sum_{t=k_{j-1}^0+1}^{k_j^0}f_tf_t^{'}\xrightarrow{p}\Sigma_F$  as $T\to \infty$ for $j=1,\cdots,m_0+1$;

(ii) There exists $d>0$ such that $\left\|\Delta_j\right\|\geq d>0$ for $j=2,\cdots,m_0+1$, where $\Delta_j =B_j\Sigma_FB_j^{'}- B_{j-1}\Sigma_FB_{j-1}^{'}$ and $B_j$ are $r\times r$ matrices.

(iii) $[B_1,\cdots,B_{m_0+1}]$ is of full row rank.
\end{assum}
\begin{assum}\label{Factor_Loadings}
$\left\| \lambda_{i,j} \right\|\leq \bar{\lambda}<\infty$ for $j=1,\cdots,m_0+1$, $i=1,\cdots,N$, and $\left\| \frac{1}{N}\Lambda^{'}\Lambda-\Sigma_{\Lambda} \right\|\rightarrow 0$ as $N\to \infty$ for some $r\times r$ positive
definite matrix $\Sigma_{\Lambda}$.
\end{assum}
\begin{assum}\label{Depen_and_Hetero}
There exists a positive constant $M<\infty$ such that

\begin{itemize}
\item[(i)] $E(e_{it})=0$ and $E|e_{it}|^8\leq M$ for all $i=1,\cdots,N$ and $t=1,\cdots,T$;
\item[(ii)] $E(e_s^{'}e_t/N)=E(N^{-1}\sum_{i=1}^Ne_{is}e_{it})=\gamma_N(s,t)$ and $\sum_{s=1}^{T}|\gamma_N(s,t)|\leq M$ for every $t\leq T$;
\item[(iii)] $E(e_{it}e_{jt})=\tau_{ij,t}$ with $|\tau_{ij,t} |<\tau_{ij}$ for some $\tau_{ij}$ and for all $t=1,\cdots,T$ and $\sum_{j=1}^{N}|\tau_{ij}|\leq M$ for every $i\leq N$;
\item[(iv)] $E(e_{it}e_{js})=\tau_{ij,ts}$ and
$\frac{1}{NT}\sum\limits_{i,j,t,s}|\tau_{ij,ts}|\leq M;
$
\item[(v)] For every $(s,t)$, $E\left| N^{-1/2}\sum_{i=1}^{N}(e_{is}e_{it}-E[e_{is}e_{it}]) \right|^4\leq M$;
\item[(vi)] For every $(i,j)$ and $\ell=1,\cdots,m_0+1$, $E\left| (k_{\ell}^0-k_{\ell-1}^0)^{-1/2}\sum_{t=k_{\ell-1}^0+1}^{k_{\ell}^0}(e_{it}e_{jt}-E[e_{it}e_{jt}]) \right|^4\leq M$.
\end{itemize}
\end{assum}
\begin{assum}\label{Weak_Dependence}

There exists a positive constant $M<\infty$ such that
\begin{eqnarray*}
E\left(\frac{1}{N}\sum\limits_{i=1}^{N}\left\| \frac{1}{\sqrt{k_j^0-k_{j-1}^0}}\sum\limits_{t=k_{j-1}^0+1}^{k_j^0}f_te_{it} \right\|^2\right)&\leq& M\,\,\,\,\text{for $j=1,\cdots,m_0+1$}.
\end{eqnarray*}
\end{assum}
\begin{assum}\label{eigenvalues}
The eigenvalues of $\Sigma_G\Sigma_\Lambda$ are distinct, and $\Sigma_G=E(g_tg_t')$.
\end{assum}
\begin{assum}\label{Hajek-Renyi}
Define $\epsilon_t=f_tf_t^{'}-\Sigma_F$. The H\'{a}jek--R\'{e}nyi inequality applies to the processes $\{\epsilon_t,t=k_{j-1}^0+1,\cdots,k_j^0\}$ and
$\{\epsilon_t,t=k_j^0,\cdots,k_{j-1}^0+1\}$ for $j=1,\cdots,m_0+1$.
\end{assum}

\begin{assum}\label{error_sup}
For a fixed $s$ and varying $k$ (or vice-versa), there exists a constant $M < \infty$ such that for all $k-s > R \equiv (\log T)^2$:

(i) For each $\ell = 1, \dots, T$:
$$\max_{k-s > R} \frac{1}{k-s} \sum_{t=s+1}^{k} \left| \frac{1}{\sqrt{N}} \sum_{i=1}^{N} \left[ e_{i\ell}e_{it} - E(e_{i\ell}e_{it}) \right] \right|^2 = O_p(1);$$

(ii) $\max_{k-s > R} \frac{1}{k-s} \sum_{t=s+1}^{k} \left\| \frac{1}{\sqrt{N}} \sum_{i=1}^{N} \lambda_i e_{it} \right\|^2 = O_p(1);$

(iii) $\max_{k-s > R} \frac{1}{k-s} \sum_{t=s+1}^{k} \left\| \frac{1}{\sqrt{NT}} \sum_{h=1}^{T} \sum_{i=1}^{N} f_h [e_{ih}e_{it} - E(e_{ih}e_{it})] \right\|^2 = O_p(1).$
\end{assum}

\begin{assum}\label{Central_Limit}
The following conditions hold for some finite constant $M > 0$:

(i) For a fixed $s$ and varying $k$ (or vice-versa), such that $k-s > (\log T)^2$:
$$\limsup_{N, (k-s) \to \infty} \left\| \frac{1}{\sqrt{N(k-s) \log\log(N(k-s))}} \sum_{t=s+1}^k \sum_{i=1}^N \lambda_i f_t' e_{it} \right\| \leq M;$$
And for any constant fraction $\eta \in (0,1]$, let $\mathcal{S}_{\eta} = \{(s,k) : k-s > \eta T\}$. Then
$$\sup_{(s,k) \in \mathcal{S}_{\eta}} \left\| \frac{1}{\sqrt{N(k-s)}} \sum_{t=s+1}^k \sum_{i=1}^N \lambda_i f_t' e_{it} \right\| = O_p(1);$$ 

(ii) For a fixed $s$ and varying $k$ (or vice-versa), such that $k-s > (\log T)^2$:
$$\limsup_{N, (k-s) \to \infty} \left\| \frac{1}{N\sqrt{k-s} \sqrt{\log\log(N^2(k-s))}} \sum_{t=s+1}^{k} \sum_{i=1}^{N} \sum_{j=1}^{N} \lambda_{i} \lambda_{j}' (e_{it} e_{jt} - E[e_{it} e_{jt}]) \right\| \leq M;$$
And for any constant fraction $\eta \in (0,1]$, let $\mathcal{S}_{\eta} = \{(s,k) : k-s > \eta T\}$. Then
$$\sup_{(s,k) \in \mathcal{S}_{\eta}} \left\| \frac{1}{N\sqrt{k-s} } \sum_{t=s+1}^{k} \sum_{i=1}^{N} \sum_{j=1}^{N} \lambda_{i} \lambda_{j}' (e_{it} e_{jt} - E[e_{it} e_{jt}]) \right\| = O_p(1).$$
\end{assum}

Assumptions \ref{factors}--\ref{eigenvalues} are standard in the factor model literature. Assumption \ref{factors} (i) is similar to Assumption A in \cite{Bai2003} and (ii) imposes restrictions on $B_j$ and $B_{j-1}$ to ensure the identification of the breakpoint.  Assumption \ref{factors} (ii) rules out the case that $B_j=-B_{j-1}$ since the objective function (\ref{obj1_fun}) cannot identify the difference between $\hat \Sigma_{j-1}$ and $\hat \Sigma_{j}$. Our analysis focuses on the case of fixed-magnitude breaks to facilitate the development of the statistical properties of near-zero eigenvalues established in Proposition 2. Assumption \ref{factors} (iii) implies that $\Sigma_G$ is positive definite. Assumption \ref{Factor_Loadings} is similar to Assumption B of \cite{Bai2003}. Assumption \ref{Depen_and_Hetero} allows for weakly correlated error terms in both time and cross-sectional dimensions. Assumption \ref{Weak_Dependence} means that the factors and idiosyncratic errors are allowed to be weakly dependent within each regime. Assumption \ref{eigenvalues} corresponds to Assumption G in \cite{Bai2003}. Assumption \ref{Hajek-Renyi} corresponds to Assumption 7 of \cite{Baltagi2021}, which allows the H\'{a}jek--R\'{e}nyi inequality to be applicable to the second moment process of the factors. 

Assumptions \ref{error_sup} and \ref{Central_Limit}
are similar to Assumptions 7
and 8 of DBH. 
The terms within the squared norm in Assumption \ref{error_sup} are standard in the factor model literature and are typically $O_p(1)$ (e.g., Assumptions C(5), F(1), and F(3) in \cite{Bai2003}). Our assumption strengthens these traditional conditions by requiring that the averages of these squared terms remain uniformly $O_p(1)$ over sub-intervals with length exceeding $(\log T)^2$. We provide sufficient conditions for this uniformity in Section C of the Supplementary Appendix.
Assumption \ref{Central_Limit} concerns the partial sums of the mean-zero processes $Z_{N,t}^{(1)}$ and $Z_{N,t}^{(2)}$ over candidate sub-intervals $(s, k]$, where $Z_{N,t}^{(1)}=\frac{1}{\sqrt N}\sum_{i=1}^{N}\lambda_i f_t'e_{it}$ and $Z_{N,t}^{(2)}=\frac{1}{N}\sum_{i=1}^{N}\sum_{j=1}^{N}\lambda_i\lambda_j'\bigl(e_{it}e_{jt}-E[e_{it}e_{jt}]\bigr)$. 
For $k-s \ge \eta T$, the required $O_p(1)$ bound is a direct consequence of a Functional Central Limit Theorem. For shorter intervals where $k-s \ge (\log T)^2$, the relevant upper envelope involves a $\log\log$ factor, consistent with the Law of the Iterated Logarithm. Detailed derivations for a benchmark example illustrating the feasibility of Assumption \ref{Central_Limit} are provided in Section C of the Supplementary Appendix.

\begin{assum}\label{as-.LeeL}
With probability approaching one (w.p.a.1), the following inequalities hold:
\begin{align*}
0<\underline{c}\le \min_{k-s\geq \sqrt{T}}\rho_{r}\left(\frac{1}{N(k-s)}\sum_{t=s+1}^{k}\Lambda^{\prime}e_{t}e_{t}^{\prime}\Lambda\right)\text{ and }
\rho_{1}\left(\frac{1}{NT}\sum_{t=1}^{T}\Lambda^{\prime}e_{t}e_{t}^{\prime}\Lambda\right)\le \overline{c}<+\infty,
\end{align*}
as $N,T\to\infty$, where $\underline{c}$ and $\overline{c}$ are some constants.
\end{assum}
 \begin{assum}\label{B_C_full_rank_project} For
$j=1,\cdots,m_{0}$, $\|B_{j}f_{k_{j}^{0}}-\mathrm{Proj}(B_{j}f_{k_{j}^{0}}|B_{j+1})\|\ge d>0$\\
when $B_{j+1}$ is singular or $\|B_{j+1}f_{k_{j}^{0}+1}-\mathrm{Proj}(B_{j+1}f_{k_{j}^{0}+1}|B_{j})\|\ge d>0$
when $B_{j}$ is singular, where $\mathrm{Proj}(\cdot|\mathbb{Z})$
denotes the projection of a vector onto the columns of $\mathbb{Z}$,
and $d$ is a constant. \end{assum}

Assumption \ref{as-.LeeL} extends Assumption
F3 in \cite{Bai2003} and ensures that $\rho_{r}(\hat{\Sigma}(k,s))\asymp\frac{1}{N}$ w.p.a.1
if the interval $[k+1,s]$ belongs to $j$-th regime and $B_{j}$ is singular.
Assumption \ref{B_C_full_rank_project} is critical when $B_j$ or $B_{j+1}$ is singular. 
Suppose $B_{j+1}$ is singular with $r_{j+1} < r$. Under the true partition, the pseudo-factor covariance matrix for $(k_j^0, k_{j+1}^0]$ has rank $r_{j+1}$. 
Shifting the split point from $k_j^0$ to $k_j^0-1$ mistakenly incorporates $B_j f_{k_j^0}$ from regime $j$ into the misaligned segment $(k_j^0-1, k_{j+1}^0]$.
Assumption \ref{B_C_full_rank_project} requires $B_j f_{k_j^0} \notin \mathrm{Col}(B_{j+1})$, ensuring that the misaligned segment's rank exceeds $r_{j+1}$. In finite samples, where estimated factors are used to compute $\hat{\Sigma}(k_{\ell-1}, k_\ell)$, this rank inflation translates into an increase in the objective function, facilitating precise breakpoint identification. Conversely, if $B_j f_{k_j^0} \in \mathrm{Col}(B_{j+1})$ (e.g., $f_{k_j^0} = 0$), misspecifying the breakpoint as $k_j^0-1$ fails to increase the rank of segment $(k_j^0-1, k_{j+1}^0]$, in which case the breakpoint cannot be precisely estimated.
The case of $B_{j}$ being singular is similar by symmetry.

\vspace{-1em}
\subsection{Asymptotic properties of the QML estimators}
\vspace{-1em}

In this subsection, we establish the asymptotic properties
of the QML estimators. The analysis of these properties is significantly
more challenging in a multi-break setting than in a single-break setting.
As discussed in the Introduction, the estimation error of any estimated
breakpoint is generally interdependent with that of others. This interdependence
is further complicated when both Type 1 and Type 2 breaks are present.
Furthermore, since the true endpoints of a regime are unknown, an
estimated segment $[\hat{k}_{j}+1,\hat{k}_{j+1}]$ may have no overlap
with the true interval $[k_{j}^{0}+1,k_{j+1}^{0}]$. This contrasts
sharply with the single-break setting, where the estimated segments
are guaranteed to overlap with the true regimes because $k_{0}^{0}=0$
and $k_{m_{0}+1}^{0}=k_{2}^{0}=T$ are known.

To address these challenges, we introduce a strategy
of ``conceptually'' overfitting the number of breaks to isolate the $j$-th breakpoint
of interest. Specifically, assuming $\hat{k}_{j}\ne k_{j}^{0}$, we
conceptually insert all true breakpoints, except for the $j$-th one,
into the existing set of estimated breakpoints. After inserting these
breakpoints, all resulting segments (derived from the estimated and
inserted breakpoints) contain no true breakpoint, except for one segment
that contains the $j$-th true breakpoint. This enables us to derive
an upper bound for $|\hat{k}_{j}-k_{j}^{0}|$.

While this strategy helps isolate the target breakpoint
and analyze its asymptotic properties, the conceptually overfitting strategy introduces
two technical challenges. First, an inserted true breakpoint may be
very close to an estimated breakpoint, leading to a segment with fewer
than $r$ observations. In such a segment, the pseudo-factors will
have an exactly singular covariance matrix. We address this issue
by merging such a short segment with an adjacent regime whose length
is necessarily proportional to $T$. More details are provided in
the text preceding the proof of Theorem 1 in the Appendix. Second, overfitting the breaks inevitably cuts a true regime into sub-regimes.
When the pseudo-factors in the $j$-th regime are not of full column
rank (i.e., $B_{j}$ is singular), it is crucial and challenging to
compare the nearly zero eigenvalues of the sample covariance matrices
of pseudo-factors in the true regimes versus those in the sub-regimes
created by conceptually overfitting. Proposition \ref{rho_difference_Sigma_Sigma1}
provides a detailed rate for the difference between these nearly-zero
eigenvalues. This proposition plays a crucial role in establishing
the point-consistency of the QML estimator and the selection consistency
of our proposed information criterion.

To compare these nearly-zero eigenvalues,
we consider the case of conceptually overfitting a point $\tilde{k}_{j}$ within
the regime $[k_{j-1}^{0}+R,\ k_{j}^{0}-R]$, where $R=(\log T)^{2}$.
Recall that $\hat{\Sigma}(k,s)$ denotes the estimated sample covariance matrix of $\hat{g}_t$ for $k+1 \leq t \leq s$.
Specifically,
\[
\hat{\Sigma}(k_{j-1}^{0},k_{j}^{0})=\hat{G}_{j}^{\prime}\hat{G}_{j}/(k_{j}^{0}-k_{j-1}^{0}),\ \ \hat{\Sigma}(k_{j-1}^{0},\tilde{k}_{j})=\hat{G}_{j,1}^{\prime}\hat{G}_{j,1}/(\tilde{k}_{j}-k_{j-1}^{0}),\ \ \hat{\Sigma}(\tilde{k}_{j},k_{j}^{0})=\hat{G}_{j,2}^{\prime}\hat{G}_{j,2}/(k_{j}^{0}-\tilde{k}_{j}),\]
 where $\hat{G}_{j}=[\hat{g}_{k_{j-1}^{0}+1},...,\hat{g}_{k_{j}^{0}}]'$,
$\hat{G}_{j,1}=[\hat{g}_{k_{j-1}^{0}+1},...,\hat{g}_{\tilde{k}_{j}}]'$,
$\hat{G}_{j,2}=[\hat{g}_{\tilde{k}_{j}+1},...,\hat{g}_{k_{j}^{0}}]'$.
The following proposition then establishes the convergence rate of
the difference between the eigenvalues of $\hat{\Sigma}(k_{j-1}^{0},k_{j}^{0})$
and $\hat{\Sigma}(k_{j-1}^{0},\tilde{k}_{j})$, which is crucial for
the subsequent theoretical analysis when $B_{j}$ is singular.

\begin{proposition} \label{rho_difference_Sigma_Sigma1} Under Assumptions \ref{factors}--\ref{Central_Limit} and $\frac{N}{T}\to \kappa$, as $N,\,T\to \infty$ for $0<\kappa<\infty$, when $\mathrm{rank}(B_{j})=r_{j}<r$, then
\begin{eqnarray*}
 &  & (i)\,\,\,\, N\sqrt{\tilde{k}_{j}-k_{j-1}^{0}}\left|\rho_{\ell}\left(\hat{\Sigma}(k_{j-1}^{0},k_{j}^{0})\right)-\rho_{\ell}\left(\hat{\Sigma}(k_{j-1}^{0},\tilde{k}_{j})\right)\right|=O_{p}\left(\log\log NT\right),\\
 &  & (ii)\,\,\,\, N\sqrt{k_{j}^{0}-\tilde{k}_{j}}\left|\rho_{\ell}\left(\hat{\Sigma}(k_{j-1}^{0},k_{j}^{0})\right)-\rho_{\ell}\left(\hat{\Sigma}(\tilde{k}_{j},k_{j}^{0})\right)\right|=O_{p}\left(\log\log NT\right)\end{eqnarray*}
for $\ell=r_{j}+1,\cdots,r$ uniformly over $\tilde{k}_j\in [k_{j-1}^{0}+R,k_{j}^{0}-R]$ with $R=(\log T)^2$.
\end{proposition}

In Proposition \ref{rho_difference_Sigma_Sigma1},
the convergence rate is determined by $N$ and the length of the subsample,
$\tilde{k}_{j}-k_{j-1}^{0}$ (or $k_{j}^{0}-\tilde{k}_{j}$). Notably,
the distance (in terms of Frobenius norm) between $\hat{\Sigma}(k_{j-1}^{0},k_{j}^{0})$
and $\hat{\Sigma}(k_{j-1}^{0},\tilde{k}_{j})$ is no smaller than
the order $T^{-1/2}$, while the small eigenvalues of $\hat{\Sigma}(k_{j-1}^{0},k_{j}^{0})$
converge to zero at a rate of $T^{-1}$. This disparity implies that
we cannot obtain the rate of $\rho_{\ell}\left(\hat{\Sigma}(k_{j-1}^{0},k_{j}^{0})\right)-\rho_{\ell}\left(\hat{\Sigma}(k_{j-1}^{0},\tilde{k}_{j})\right)$
for $\ell=r_{j}+1,\cdots,r$ using the continuous mapping theorem.
This is because the distance between these two matrices is actually
much larger than the magnitude of their smallest eigenvalues. To address
this, Lemmas 4--6 in the Appendix develop essential technical tools
that facilitate the proof of the detailed rate for $\rho_{\ell}\left(\hat{\Sigma}(k_{j-1}^{0},k_{j}^{0})\right)-\rho_{\ell}\left(\hat{\Sigma}(k_{j-1}^{0},\tilde{k}_{j})\right)$
as established in Proposition \ref{rho_difference_Sigma_Sigma1}.
These lemmas may also be of independent interest.

\begin{theorem}Let $\epsilon\in (0,1/2)$ be a constant. Under Assumptions \ref{factors}--\ref{as-.LeeL}
and as $N,T\to\infty$ with $N/T\to\kappa$ for some $0<\kappa<\infty$:\\
(i) when $r_{j,j+1}>\min\{r_{j},r_{j+1}\}$,
$Prob(|\hat{k}_{j}-k_{j}^{0}|\ge\sqrt{T})\to0$;\\
(ii) when $r_{j,j+1}=\min\{r_{j},r_{j+1}\}$,
$Prob(|\hat{k}_{j}-k_{j}^{0}|\ge T^{\alpha})\to0$,
where $\alpha=1/2+\epsilon$\\
for $j=1,\cdots,m_{0}$. \label{fraction_consistency}
\end{theorem}

Theorem \ref{fraction_consistency} provides an
initial upper bound on the estimation error of $\hat{k}_{j}$ for
both types of breaks. Part (i) indicates that the estimation error
for Type 1 breaks is bounded by $\sqrt{T}$ with high probability
in large samples. Moreover, Part (ii) shows that for a Type 2 break,
the upper bound is $T^{\alpha}$, where $\alpha$ is some constant
slightly larger than $1/2$. These results collectively imply fraction-consistency,
i.e., $\hat \tau_{j}-\tau_{j}^{0}=o_{p}(1)$, where $\hat \tau_{j}=\hat{k}_{j}/T$
is the estimated break fraction.
Building upon Theorem \ref{fraction_consistency},
the subsequent theorem further refines the convergence rate of $\hat{k}_{j}-k_{j}^{0}$.

\begin{theorem} Under Assumptions
\ref{factors}--\ref{as-.LeeL} and $\frac{N}{T}\to\kappa$, as $N,T\to\infty$
for $0<\kappa<\infty$, \\
 (i) when $r_{j,j+1}>\min\{r_{j},r_{j+1}\}$
and Assumption \ref{B_C_full_rank_project} holds, $Prob(\hat{k}_{j}=k_{j}^{0})\to1$;\\
 (ii) when $r_{j,j+1}=\min\{r_{j},r_{j+1}\}$, $\hat{k}_{j}-k_{j}^{0}=O_{p}(1)$\\
 for $j=1,\cdots,m_{0}$, where $r_{j,j+1}=rank([B_{j},B_{j+1}])$.
\label{consistency} \end{theorem}

Theorem \ref{consistency} (i) demonstrates that the estimated
change points converge to the true breakpoints w.p.a.1 for a Type
1 break. This result is stronger than that of \cite{Baltagi2021},
who established that the distance between estimated and true breakpoints
is $O_{p}(1)$ for Type 1 changes. It is important to note that Type 1 breaks incorporate scenarios of emergence and disappearance of factors, which correspond to $r_{j,j+1}=\max\{r_{j},r_{j+1}\}$. Our QML estimator
proves to be consistent when the number of factors changes across
regimes, whereas \cite{Ma2018} excluded this case by assumption.
Furthermore, Theorem \ref{consistency} (ii) suggests that the discrepancy
between the QML estimators and the true change points is stochastically
bounded for Type 2 changes.

\begin{remark}
The core intuition of Theorem \ref{consistency} (i) lies in how the QML objective function exploits cross-sectional information via the spectral properties of the pseudo-factors.  
Misidentifying a Type 1 breakpoint incorporates ``contaminated'' data from an adjacent regime, which inflates the near-zero eigenvalues of $\hat{\Sigma}(\cdot)$.
Because the log-determinant is highly sensitive to small changes near singularity, this misalignment triggers a sharp increase in the objective function by an amount that increases with $N$. 
Hence, as $N$ grows, the objective function is minimized only at $\hat{k}_j = k_j^0$, ensuring point-consistency under Type 1 breaks.
By contrast, for Type 2 breaks the loading spaces before and after the breakpoint
coincide, so misplacing the breakpoint mainly redistributes the spectral mass among the large
eigenvalues rather than inflating the near-zero ones. Hence, the objective function does not
benefit from the near-singularity signal, and the estimator attains only the standard
$O_p(1)$ rate in Theorem 2(ii), rather than point-consistency. 
\end{remark}

\begin{remark}[Determining the types of changes]\label{rem:type_change}
We discuss how to analyze the nature of changes around the breakpoints
and assess the reliability of the estimated breakpoint.
To implement the classification rule below in practice, the ranks
$(r_j,r_{j+1},r_{j,j+1})$ can be estimated from the data using the
step-by-step procedure described in Remark~\ref{rem:rank_estimation}.
(A) If $r_j=r_{j+1}=r_{j,j+1}$, we classify the breakpoint as a rotational change (Type 2).
This includes both the full-rank case $r_{j,j+1}=r$ and the case $r_{j,j+1}<r$ (see Remark \ref{rotational_singular_rmk}).
(B) If $r_{j,j+1}>\min\{r_j,r_{j+1}\}$, the $j$-th break is a singular change (Type 1).
If $r_{j,j+1}=r_j+r_{j+1}$, the loading spaces before and after the breakpoint are linearly
independent. If $r_{j,j+1}=r_j>r_{j+1}$ (or $r_{j,j+1}=r_{j+1}>r_j$), the loading space after
(before) the breakpoint is a subspace of that before (after) the breakpoint, implying that
some factors disappear (or new factors emerge) after the breakpoint.
For case (A), Theorem~2(ii) shows that the estimation error is stochastically bounded.
For case (B), Theorem~2(i) shows that the estimated breakpoints equal the true breakpoints
w.p.a.1. Therefore, one can determine the type of break and assess the reliability of the
estimated breakpoint based on the values of $r$, $r_j$, $r_{j+1}$, and $r_{j,j+1}$.
\end{remark}
\begin{remark}[Estimating $(r_j,r_{j+1},r_{j,j+1})$ in practice]\label{rem:rank_estimation}
The number of pseudo-factors in full sample, r, is consistently estimated using the information criteria (IC) of Bai and Ng (2002). Given $\hat{r}$, we obtain the estimated breakpoints $\{\hat{k}_{1}, \dots, \hat{k}_{m_{0}}\}$. To implement the classification rule in Remark~\ref{rem:type_change}, we apply the Bai and Ng (2002) IC---with $\hat{r}$ as the upper bound---to three specific subsamples: $[\hat{k}_{j-1}+1, \hat{k}_{j}]$, $[\hat{k}_{j}+1, \hat{k}_{j+1}]$, and $[\hat{k}_{j-1}+1, \hat{k}_{j+1}]$. Let the resulting rank estimates be $\hat{r}_{j}$, $\hat{r}_{j+1}$, and $\hat{r}_{j,j+1}$, respectively. 
Theorem 2 establishes that $\hat{k}_j$ is point-consistent under Type 1 changes and satisfies $\hat{k}_{j}-k_{j}^{0}=O_{p}(1)$ under Type 2 changes. Since the distance between true breakpoints is of order $T$, the $O_p(1)$ estimation error in the breakpoints has a negligible effect on rank estimation; thus, Bai and Ng's IC remains consistent for these subsamples. 
To ensure finite-sample robustness, we apply the adjustment $\tilde{r}_{j,j+1} = \max\{\hat{r}_{j,j+1}, \hat{r}_j, \hat{r}_{j+1}\}$, which guarantees $\max\{ \hat{r}_j, \hat{r}_{j+1}\} \le \tilde{r}_{j,j+1} \le \hat{r}$ by construction. The breakpoint type is then determined via $(\hat{r}_j, \hat{r}_{j+1}, \tilde{r}_{j,j+1})$ according to Remark~\ref{rem:type_change}.
\end{remark}

{\bf Determining the number of breaks}

We now discuss the choice of the number of breaks $m$, which
is an important issue when the objective function is used in practice. 
We propose selecting $m$ to
minimize the following information criterion: 
\begin{eqnarray}
IC(m) & = & \sum\limits _{\ell=1}^{m+1}(\hat k_{\ell}-\hat k_{\ell-1})\log|\hat{\Sigma}(\hat k_{\ell-1},\hat k_{\ell})|+m(1+|\hat{\rho}|)r^{2}\log(\min(N,T)),\label{model_selection}\end{eqnarray}
where $(\hat{k}_1, \dots, \hat{k}_m)$ are obtained by solving (\ref{obj1_fun}) for each given $m$, and $\hat{\rho}$ is the radius of the estimated AR(1) coefficient
matrix, obtained by fitting a VAR(1) model to $\hat{g}_{t}$, although the actual $g_t$ is not required to be an AR(1) process. 

Note that, for any fixed $m$, the penalty term in Eq.~(\ref{model_selection}) is constant with respect to the breakpoint locations. Hence, the joint search over $m$ and the breakpoint locations is numerically equivalent to the following two-step procedure: first, for each $m$, determine the optimal break locations by minimizing the QML objective function in Eq.~(\ref{obj1_fun}); second, substitute these minimizers into Eq.~(\ref{model_selection}) and select the value of $m$ that yields the smallest criterion value. In practice, for each candidate $m$, the minimization of $U_{NT}$ is implemented by the dynamic programming algorithm in Section~2.2. 

Let
$m_{max}$ be a bounded integer such that $m_{0}<m_{max}$ and $\hat{m}=\arg\min_{0\leq m\leq m_{max}}IC(m)$.
\begin{theorem}\label{select_consistency} Under the Assumptions of Theorem 2,
$$
\lim_{N,T\rightarrow\infty}P(\hat{m}=m_{0})=1.$$\end{theorem}

\begin{remark}\label{joint_consistency}
Theorems \ref{consistency} and \ref{select_consistency} together imply the following joint consistency result:
\[
P(\hat m=m_0,\ \hat k_j=k_j^0,\ j\in S)\to 1 \quad \text{as } N,T\to\infty,
\]
where
$S=\{j \mid r_{j,j+1}>\min\{r_j,r_{j+1}\},\ j=1,\ldots,m_0\}$
denotes the set of indices for singular breaks. 
If we also
include rotational changes, then for any given $\epsilon>0$ there exists an $M>0$ such that
$P( |\hat{k}_j - k_{j}^0| < M,\, j=1,2,\cdots,m_0,\   \hat{m} = m_0)>1-\epsilon$.
\end{remark}
\begin{remark}\label{select_consistency_remark}
The penalty in (\ref{model_selection}) does not differentiate the break types; instead, it is designed to ensure the consistency of $\hat{m}$ by leveraging the different divergence rates of the objective function when $m$ is misspecified, as detailed in the proof of Theorem~\ref{select_consistency}. Specifically, when $m>m_{0}$, overfitting changes the objective function by only $O_{p}(1)$ regardless of the break type, so the penalty term of order $\log(\min(N,T))$ asymptotically dominates this gain and rules out overfitting. When $m<m_{0}$, underfitting increases the objective function by order $T\log N$ if the omitted break is Type~1 and by order $T$ if it is Type~2; both dominate the penalty term, so the criterion does not favor underfitting. Thus, the same penalty form is sufficient for both break types. The factor $r^{2}$ reflects model complexity through the dimension of the factor covariance matrix, and $(1+|\hat{\rho}|)$ is a finite-sample adjustment that prevents overfitting when $\hat{g}_{t}$ exhibits stronger serial dependence. Importantly, it is not meant to impose, nor does it require, that the pseudo-factors $g_{t}$ follow a VAR(1) model.
\end{remark}

\vspace{-1em}
\section{Simulation}
\vspace{-1em}

In this section, we consider DGPs corresponding to different break
types to evaluate the finite sample performance of the QML estimator.
We compare the QML estimator with two other estimators. As shown below,
$\hat{k}_{BKW}$ is the estimator proposed by \citetalias{Baltagi2021}
(\citeyear{Baltagi2021}, BKW hereafter); $\hat{k}_{MS}$ is the estimator proposed by
\citeauthor{Ma2018} (\citeyear{Ma2018}, MS hereafter); and $\hat{k}_{QML}$ is the QML estimator.
We explore the performance of these estimators \footnote{We also evaluate the estimators of \cite{Ma2023}, which perform similarly to those of \cite{Ma2018}. Due to space constraints, a comprehensive comparison of these results is provided in Appendix D of the supplementary material.}
under various DGPs, incorporating different types of changes. We
calculate the RMSE of these change point estimators $\hat{k}_{BKW}$, $\hat{k}_{MS}$,
and $\hat{k}_{QML}$, and each experiment is repeated 1000 times,
where RMSE$_{i}=\sqrt{\frac{1}{1000}\sum\limits _{s=1}^{1000}(\hat{k}_{i,s}-k_{i}^{0})^{2}}$
for $i=1,\cdots,m_{0}$. Note that the MS method sometimes detects
more or fewer than $m_{0}$ breaks. To ensure comparability, the RMSE for the MS estimator is calculated exclusively from
instances where the method correctly identifies the true number of
breaks.

Each factor is generated by the following AR(1) process: \begin{eqnarray}
f_{tp}=\rho f_{t-1,p}+u_{t,p},\quad for\quad t=2,\cdots,T;\quad p=1,\cdots,r_{0},\label{eq:dgp factor}\end{eqnarray}
 where $u_{t}=(u_{t,1},\cdots,u_{t,r_{0}})^{'}$ is i.i.d. $N(0,I_{r_{0}})$
for $t=2,\cdots,T$ and $f_{1}=(f_{1,1},\cdots,f_{1,r_{0}})^{'}$
is drawn from $N(0,\frac{1}{1-\rho^{2}}I_{r_{0}})$. The scalar $\rho$
captures the serial correlation of factors, and the idiosyncratic
errors are generated by \begin{eqnarray*}
e_{i,t}=\alpha e_{i,t-1}+v_{i,t},\quad for\quad i=1,\cdots,N;\quad t=2,\cdots,T,\end{eqnarray*}
 where $v_{t}=(v_{1,t},\cdots,v_{N,t})^{'}$ is i.i.d. $N(0,\Omega)$
for $t=2,\cdots,T$ and $e_{1}=(e_{1,1},\cdots,e_{N,1})^{'}$ is $N(0,\frac{1}{(1-\alpha^{2})}\Omega)$.
The scalar $\alpha$ captures the serial correlation of the idiosyncratic
errors, and $\Omega$ is generated as $\Omega_{ij}=\beta^{|i-j|}$
so that $\beta$ captures the degree of cross-sectional dependence
of the idiosyncratic errors. In addition, $u_{t}$ and $v_{t}$ are
mutually independent for all values of $t$. We consider the following
DGPs for factor loadings and investigate the performance of the QML
estimator for the different types of breaks discussed in Section 3.

\textbf{DGP 1.} We first consider the case in which $m_0=2$ and $r_0=3$. We set $k_1^0=[0.3T]$ and $k_2^0=[0.7T]$, where $[\cdot]$ denotes the rounding operation. This means that there are two breakpoints and three regimes.

\textbf{DGP 1.A}
In this setup, we adopt the DGP from MS. Each of the three
regimes has $r_{0}$ effective factors, i.e., $r_{1}=r_{2}=r_{3}=3$.
The factor loading matrices on the effective factors are generated
as follows:
\[
\breve{\Lambda}_{1}=(\breve{\lambda}_{1,1},\ldots,\breve{\lambda}_{N,1})',\quad\breve{\lambda}_{i,1}\;\overset{\text{i.i.d.}}{\sim}\; N\bigl((0.5b,\,0.5b,\,0.5b)^{\prime},\,\tfrac{1}{r_{0}}I_{r_{0}}\bigr),\]
 \[
\breve{\Lambda}_{2}=(\breve{\lambda}_{1,2},\ldots,\breve{\lambda}_{N,2})',\quad\breve{\lambda}_{i,2}\;\overset{\text{i.i.d.}}{\sim}\; N\bigl((b,\, b,\, b)^{\prime},\,\tfrac{1}{r_{0}}I_{r_{0}}\bigr),\]
 \[
\breve{\Lambda}_{3}=(\breve{\lambda}_{1,3},\ldots,\breve{\lambda}_{N,3})',\quad\breve{\lambda}_{i,3}\;\overset{\text{i.i.d.}}{\sim}\; N\bigl((1.5b,\,1.5b,\,1.5b)^{\prime},\,\tfrac{1}{r_{0}}I_{r_{0}}\bigr),\]
The $\Lambda$ matrix in (\ref{Baltagi}) can be written as $\Lambda\;=\;\bigl[\breve{\Lambda}_{1},\breve{\Lambda}_{2},\breve{\Lambda}_{3}\bigr]_{N\times3r_{0}}$.
We define the three $3r_{0}\times 3r_{0}$ block-selector matrices\[
B_{1}=\operatorname{diag}\!\bigl(I_{r_{0}},\,0_{r_{0}\times r_{0}},\,0_{r_{0}\times r_{0}}\bigr),\quad
B_{2}=\operatorname{diag}\!\bigl(0_{r_{0}\times r_{0}},\,I_{r_{0}},\,0_{r_{0}\times r_{0}}\bigr),\quad
B_{3}=\operatorname{diag}\!\bigl(0_{r_{0}\times r_{0}},\,0_{r_{0}\times r_{0}},\,I_{r_{0}}\bigr),
\]
 so that in regime $j$ we set $\Lambda_{j}=\Lambda\, B_{j}$, 
which yields three effective factors per regime and a total of $r=3r_{0}=9$
pseudo-factors over the full sample. This corresponds to a Type 1
breakpoint scenario with $r_{j,j+1}=r_{j}+r_{j+1}$,
where loadings in adjacent regimes are independent. By
varying $b$ we can control the magnitude of the break.

\textbf{DGP 1.B} We set $\lambda_{i}$ to be i.i.d.
$N(0,\frac{1}{r}I_{r})$ across $i$, and define $\Lambda=(\lambda_{1},\ldots,\lambda_{N})'$.
In the first regime, $\Lambda_{1}=(\lambda_{1,1},\cdots,\lambda_{N,1})^{'}=\Lambda B_{1}$
with $B_{1}=[1,0,0;0,1,0;0,0,0]$. In the second regime, $\Lambda_{2}=(\lambda_{1,2},\cdots,\lambda_{N,2})^{'}=\Lambda B_{2}$
with $B_{2}=[1,0,0;0,0,0;0,0,1]$. In the third regime, $\Lambda_{3}=(\lambda_{1,3},\cdots,\lambda_{N,3})^{'}=\Lambda B_{3}$
with $B_{3}=[0,0,0;0,1,0;0,0,1]$. For the full sample, the number
of pseudo-factors is $r=3$. In each of the three regimes, the number
of effective factors is 2, as $\text{rank}(B_{1})=\text{rank}(B_{2})=\text{rank}(B_{3})=2$.
This setting corresponds to a Type 1 break in Section 3 with $r_{j,j+1}>\max\{r_{j},r_{j+1}\}$.

\textbf{DGP 1.C}
The generation of $\Lambda$ remains the same as in \textbf{DGP 1.B}. For $B_j$, we set
$B_{1}=[1,0,0;0,1,0;0,0,1]$, $B_{2}=[1,0,0;0,1,0;0,0,0]$ and $B_{3}=[0,0,0;0,0,0;0,0,1]$. The number of pseudo-factors for
the full sample is $r=3$, with the numbers of effective
factors in the three regimes being 3, 2, and 1, respectively.
In the first regime, all three factors are present. In the second
regime, only the first two factors are present and in the third regime, only the third factor is present.

\textbf{DGP 1.D} The generation of $\Lambda$ remains the same as in \textbf{DGP 1.B}. Let $\Lambda_{1}=\Lambda$, $\Lambda_{2}=2\Lambda$
and $\Lambda_{3}=\Lambda$. In this case,
the number of factors remains constant across regimes, and the changes
in factor loading are purely proportional, corresponding to a setting
with two Type 2 breaks.

\textbf{DGP 1.E} In this
setup, we consider a case involving both types of changes. The generation of $\Lambda$ remains the same as in \textbf{DGP 1.B}.
We set $B_{1}=[1,0,0;0,1,0;0,0,0]$, $B_{2}=[2,x,y;0,2,z;0,0,0]$ where $x$, $y$, $z$ are i.i.d.
$N(0,1)$ and $B_{3}=[0,0,0;0,0,0;0,0,1]$.
By construction, there exists a nonsingular matrix
${\bf R}$ such that $B_{1}{\bf R}=B_{2}$, so the first breakpoint is Type~2 with
$r=3>r_{1,2}=r_{1}=r_{2}=2$.
The second breakpoint, however, is a Type 1 change and is responsible
for the rank deficiency of the pseudo-factors in the first two regimes.

Since the computation of $\hat{k}_{MS}$
requires the number of effective factors ($r_{0}$), we use $\hat{r}=r_{0}$
for it. Moreover, as $\hat{k}_{QML}$ and $\hat{k}_{BKW}$
require the number of pseudo-factors ($r$), we set $\hat{r}=r$ for
them. This allows us to focus on the estimation errors of
the breakpoint estimators themselves, rather than errors resulting
from misspecifying the number of factors.

Table~\ref{DGP1} reports the RMSEs of the estimated break dates under DGP~1.A--1.E for $(\rho,\alpha,\beta)=(0.7,0.3,0.3)$.
For DGP~\ref{DGP1}.A (both $b=1$ and $b=0$), the QML estimator yields very small RMSE in small samples, and the RMSE quickly decreases as $(N,T)$ increase; BKW and MS also improve with sample size but are uniformly less accurate than QML. 
In DGP~\ref{DGP1}.B, QML again achieves the smallest RMSE and continues to improve with $(N,T)$, whereas BKW exhibits non-vanishing (stochastically bounded) errors as the sample size grows; MS improves but remains less accurate than QML. 
In DGP~\ref{DGP1}.C, where factors disappear and the factor dimension changes across regimes, both competing methods deteriorate---especially MS, which relies on a constant factor dimension---while QML remains stable and its RMSE decreases with $(N,T)$. 
For DGP~\ref{DGP1}.D (rotational change), QML delivers small and bounded estimation errors, consistent with Theorem~\ref{consistency}(ii), and it clearly outperforms BKW; MS again performs poorly and does not improve with sample size. 
Finally, for the mixed design in DGP~\ref{DGP1}.E (Type~2 followed by Type~1), QML yields bounded RMSE for the Type~2 breakpoint and near-zero RMSE for the Type~1 breakpoint, whereas both BKW and MS have substantially larger errors. 
Overall, the simulation evidence is consistent with Theorem~\ref{consistency}: QML is consistent under Type~1 changes and attains $O_p(1)$ estimation errors under Type~2 changes. Additional results for other values of $(\rho,\alpha,\beta)$ are reported in Appendix~D of the supplementary material.

\begin{table}[!t]
\centering
\scriptsize
\setlength{\tabcolsep}{1.0em} 
\renewcommand{\arraystretch}{0.5} 
\captionsetup{skip=2pt}
\caption{Simulated root mean squared errors (RMSEs) of $\hat{k}_{QML}$, $\hat{k}_{BKW}$, and $\hat{k}_{MS}$ under DGPs 1.A--1.E}
\label{DGP1}
\begin{tabular}{l c c c c}
\toprule
& & \textbf{QML} & \textbf{BKW} & \textbf{MS} \\
\cmidrule(lr){3-3} \cmidrule(lr){4-4} \cmidrule(lr){5-5}
\textbf{DGP} & \textbf{($N,T$)} & $(\hat{k}_1,\hat{k}_2)$ & $(\hat{k}_1,\hat{k}_2)$ & $(\hat{k}_1,\hat{k}_2)$ \\
\midrule

\multicolumn{5}{l}{\textbf{DGP 1.A} ($b=1$)} \\
\midrule
& 100,100 & \textbf{(0.118, 0.095)} & (2.471, 3.445) & (2.121, 5.266) \\
& 100,300 & \textbf{(0.055, 0.055)} & (2.317, 1.338) & (0.744, 0.532) \\
& 300,300 & \textbf{(0.000, 0.055)} & (1.167, 1.200) & (0.969, 0.693) \\
& 300,600 & \textbf{(0.045, 0.032)} & (1.046, 1.051) & (0.985, 0.823) \\
& 600,600 & \textbf{(0.032, 0.032)} & (0.818, 1.030) & (1.044, 0.870) \\
\addlinespace 

\multicolumn{5}{l}{\textbf{DGP 1.A} ($b=0$)} \\
\midrule
& 100,100 & \textbf{(0.100, 0.114)} & (3.115, 2.989) & (4.336, 1.265) \\
& 100,300 & \textbf{(0.089, 0.141)} & (1.159, 0.967) & (0.933, 0.880) \\
& 300,300 & \textbf{(0.000, 0.055)} & (1.099, 2.137) & (1.153, 0.920) \\
& 300,600 & \textbf{(0.032, 0.032)} & (0.921, 1.050) & (1.298, 0.943) \\
& 600,600 & \textbf{(0.032, 0.000)} & (0.788, 0.980) & (1.429, 0.916) \\
\addlinespace

\multicolumn{5}{l}{\textbf{DGP 1.B}} \\
\midrule
& 100,100 & \textbf{(0.531, 0.577)} & (21.127, 19.862) & (12.497, 13.936) \\
& 100,300 & \textbf{(0.504, 0.427)} & (30.374, 30.487) & (31.692, 31.232) \\
& 300,300 & \textbf{(0.344, 0.336)} & (30.739, 26.776) & (23.455, 21.124) \\
& 300,600 & \textbf{(0.295, 0.321)} & (16.831, 17.168) & (5.083, 6.411) \\
& 600,600 & \textbf{(0.221, 0.307)} & (15.492, 18.497) & (3.111, 3.017) \\
\addlinespace

\multicolumn{5}{l}{\textbf{DGP 1.C}} \\
\midrule
& 100,100 & \textbf{(1.097, 0.249)} & (17.995, 21.305) & (14.184, 3.582) \\
& 100,300 & \textbf{(0.982, 0.278)} & (35.058, 24.497) & (32.378, 24.107) \\
& 300,300 & \textbf{(0.602, 0.138)} & (31.680, 20.713) & (33.039, 2.012) \\
& 300,600 & \textbf{(0.518, 0.127)} & (35.115, 8.706) & (53.368, 1.914) \\
& 600,600 & \textbf{(0.486, 0.123)} & (31.109, 9.982) & (59.029, 2.181) \\
\addlinespace

\multicolumn{5}{l}{\textbf{DGP 1.D}} \\
\midrule
& 100,100 & \textbf{(8.565, 8.408)} & (15.741, 16.328) & (16.173, 16.614) \\
& 100,300 & \textbf{(5.619, 6.389)} & (28.825, 30.149) & (45.309, 44.115) \\
& 300,300 & \textbf{(6.326, 5.310)} & (29.341, 28.247) & (41.986, 44.212) \\
& 300,600 & \textbf{(4.921, 5.171)} & (22.760, 22.752) & (93.980, 84.635) \\
& 600,600 & \textbf{(5.691, 5.362)} & (19.944, 19.463) & (82.489, 86.675) \\
\addlinespace

\multicolumn{5}{l}{\textbf{DGP 1.E}} \\
\midrule
& 100,100 & \textbf{(7.263, 1.070)} & (25.069, 9.428) & (21.130, 7.205) \\
& 100,300 & \textbf{(5.830, 0.182)} & (44.287, 14.081) & (38.584, 18.216) \\
& 300,300 & \textbf{(6.004, 0.063)} & (41.843, 12.271) & (43.257, 3.377) \\
& 300,600 & \textbf{(5.743, 0.045)} & (34.972, 9.347) & (72.697, 1.046) \\
& 600,600 & \textbf{(5.904, 0.000)} & (30.896, 9.040) & (73.199, 2.969) \\
\bottomrule
\end{tabular}
\end{table}

Finally, we assess the precision of our information
criterion (\ref{model_selection}) in selecting the number of breaks under DGPs 1.A--E and 2 (with $m_{max}=5$).
Table \ref{determine_breaks} reports the percentage of correct
detection of the number of breaks under DGP 1.A--E (all featuring two breaks) using the information
criterion (\ref{model_selection}).
It can be seen that the proposed
information criterion can approximately correctly identify the number of
breakpoints as $N$ and $T$ increase.

\textbf{DGP 2.}
In addition, DGP 2 considers scenarios where the true number of
breaks, $m_{0}$, ranges from 0 to 4. The factors are generated by
(\ref{eq:dgp factor}) and we set $r_{0}=2$. For each regime, the
$N\times2$ factor loading matrix has elements independently drawn
from $N(0,1/r_{0})$. This implies that the number of pseudo-factors,
$r$, is $(m_{0}+1)r_{0}$.

Before estimating the factors and factor
loadings, we use \citeauthor{Bai2002}'s (\citeyear{Bai2002}) information criterion to determine the number of
factors for the full samples with $r_{max}=12$. Notably, if no breakpoint
exists, the criterion function in (\ref{model_selection}) is simply
zero, as the sample covariance of the factors becomes an identity
matrix. As Table \ref{determine_breaks01234} illustrates, the probability
of our proposed information criterion accurately selecting the true
number of breakpoints approaches one as the sample size increases.
This finding further validates our Theorem \ref{select_consistency}.

\setlength{\textfloatsep}{6pt}
\setlength{\floatsep}{4pt}
\setlength{\intextsep}{6pt}

\begin{table}[!t]
\renewcommand{\arraystretch}{0.50}
\caption{Percentage of correct detection of the number of breaks under DGP 1.A, DGP 1.B, DGP 1.C, DGP 1.D, DGP 1.E}
\centering

\setlength{\tabcolsep}{2.5pt}
\captionsetup{skip=2pt}
\scriptsize

\label{determine_breaks}
\begin{tabular}{lllllllllll} \toprule
$N,T$ & DGP 1.A $(b=1)$& DGP 1.A $(b=0)$&    DGP 1.B& DGP 1.C& DGP 1.D& DGP 1.E  & \\ \midrule
&  \multicolumn{2}{c}{$\rho=0$} & \multicolumn{2}{c}{$\alpha=0$} & \multicolumn{2}{c}{$\beta=0$} & \\ \midrule
100,100   &0.725 &0.488&1.000 &0.991 &0.331 &0.422 \\
100,300   &1.000 &1.000&1.000 &1.000 &1.000 &1.000 \\
300,300   &1.000 &1.000&1.000 &1.000 &1.000 &0.999 \\
300,600   &1.000 &1.000&1.000 &1.000 &1.000 &1.000 \\
600,600   &1.000 &1.000&1.000 &1.000 &1.000 &1.000 \\\midrule

&  \multicolumn{2}{c}{$\rho=0.7$} & \multicolumn{2}{c}{$\alpha=0$} & \multicolumn{2}{c}{$\beta=0$}  &\\ \midrule
100,100   &0.110 &0.032&1.000 &0.970 &0.075 &0.109\\
100,300   &1.000 &1.000&0.999 &0.999 &0.990 &0.956\\
300,300   &1.000 &1.000&1.000 &1.000 &0.940 &0.890\\
300,600   &1.000 &1.000&1.000 &1.000 &1.000 &1.000\\
600,600   &1.000 &1.000&1.000 &1.000 &1.000 &0.999\\\midrule

&  \multicolumn{2}{c}{$\rho=0$} & \multicolumn{2}{c}{$\alpha=0.3$} & \multicolumn{2}{c}{$\beta=0$}  &\\ \midrule
100,100   &0.634 &0.451&1.000 &0.971 &0.304 &0.360\\
100,300   &1.000 &1.000&1.000 &1.000 &1.000 &1.000\\
300,300   &1.000 &1.000&1.000 &1.000 &1.000 &0.999\\
300,600   &1.000 &1.000&1.000 &1.000 &1.000 &1.000\\
600,600   &1.000 &1.000&1.000 &1.000 &1.000 &1.000\\\midrule

&  \multicolumn{2}{c}{$\rho=0$} & \multicolumn{2}{c}{$\alpha=0$} & \multicolumn{2}{c}{$\beta=0.3$}  &\\ \midrule
100,100   &0.644 &0.391&1.000 &0.995 &0.321 &0.376\\
100,300   &1.000 &1.000&1.000 &1.000 &1.000 &1.000\\
300,300   &1.000 &1.000&1.000 &1.000 &1.000 &0.999\\
300,600   &1.000 &1.000&1.000 &1.000 &1.000 &1.000\\
600,600   &1.000 &1.000&1.000 &1.000 &1.000 &1.000\\\midrule

&  \multicolumn{2}{c}{$\rho=0.7$} & \multicolumn{2}{c}{$\alpha=0.3$} & \multicolumn{2}{c}{$\beta=0.3$}  &\\ \midrule
100,100   &0.052 &0.012&1.000 &0.956 &0.069 &0.122\\
100,300   &1.000 &1.000&1.000 &1.000 &0.984 &0.958\\
300,300   &1.000 &1.000&1.000 &1.000 &0.930 &0.899\\
300,600   &1.000 &1.000&1.000 &1.000 &1.000 &1.000\\
600,600   &1.000 &1.000&1.000 &1.000 &1.000 &0.999\\\midrule

\end{tabular}
\end{table}

\vspace{-0.6em}

\begin{table}[!t]
\renewcommand{\arraystretch}{0.50}
\caption{Percentage of correct detection of the number of breaks for different $m_0$.}
\centering

\setlength{\tabcolsep}{2.5pt}
\captionsetup{skip=2pt}
\scriptsize

\label{determine_breaks01234}
\begin{tabularx}{\textwidth}{l*{5}{X}} \toprule
$N,T$ & $m_0=$0 & $m_0=$1 & $m_0=$2 & $m_0=$3 & $m_0=$4 \\ \midrule
\multicolumn{6}{c}{$\rho=0,\alpha=0,\beta=0$} \\ \midrule
100,300 &1.000 &1.000 &1.000 &1.000 &0.931  \\
300,300 &1.000 &1.000 &1.000 &1.000 &1.000  \\
100,600 &0.980 &1.000 &1.000 &1.000 &1.000   \\
300,600 &1.000 &1.000 &1.000 &1.000 &1.000 \\
600,600 &1.000 &1.000 &1.000 &1.000 &1.000\\ \midrule
\multicolumn{6}{c}{$\rho=0.7,\alpha=0,\beta=0$} \\ \midrule
100,300 &0.920 &1.000 &1.000 &1.000 &0.105  \\
300,300 &0.980 &1.000 &1.000 &1.000 &0.622 \\
100,600 &0.880 &1.000 &1.000 &1.000 &1.000 \\
300,600 &1.000 &1.000 &1.000 &1.000 &1.000 \\
600,600 &0.960 &1.000 &1.000 &1.000 &1.000 \\ \midrule
\multicolumn{6}{c}{$\rho=0,\alpha=0.3,\beta=0$} \\ \midrule
100,300 &1.000 &1.000 &1.000 &1.000 &0.845  \\
300,300 &1.000 &1.000 &1.000 &1.000 &0.999  \\
100,600 &0.980 &1.000 &1.000 &1.000 &1.000\\
300,600 &1.000 &1.000 &1.000 &1.000 &1.000 \\
600,600 &1.000 &1.000 &1.000 &1.000 &1.000 \\ \midrule
\multicolumn{6}{c}{$\rho=0,\alpha=0,\beta=0.3$} \\ \midrule
100,300 &1.000 &1.000 &1.000 &1.000 &0.904  \\
300,300 &1.000 &1.000 &1.000 &1.000 &1.000  \\
100,600 &0.980 &1.000 &1.000 &1.000 &1.000 \\
300,600 &1.000 &1.000 &1.000 &1.000 &1.000 \\
600,600 &1.000 &1.000 &1.000 &1.000 &1.000 \\ \midrule
\multicolumn{6}{c}{$\rho=0.7,\alpha=0.3,\beta=0.3$} \\ \midrule
100,300 &0.920 &1.000 &1.000 &1.000 &0.029  \\
300,300 &0.980 &1.000 &1.000 &1.000 &0.449  \\
100,600 &0.980 &1.000 &1.000 &1.000 &1.000  \\
300,600 &0.940 &1.000 &1.000 &1.000 &1.000 \\
600,600 &1.000 &1.000 &1.000 &1.000 &1.000 \\ \midrule
\end{tabularx}
\end{table}

\vspace{-1em}
\section{Empirical application}
\vspace{-1em}

In this section we apply our method to detect breakpoints
for the FRED-MD (Federal Reserve Economic Data - Monthly Data) data
set. The FRED database is maintained by the research division of the
Federal Reserve Bank of St. Louis, and is publicly accessible and
updated in real time. Our analysis utilizes
a sample of 126 unbalanced monthly time series, covering the period
from January 1959 to July 2024. These series are categorized
into eight groups: (1) output and income, (2) labor market, (3) housing,
(4) consumption, orders, and inventories, (5) money and credit, (6)
interest rates and exchange rates, (7) prices, and (8) stock market.
Following the literature, we replace the missing values using the expectation-maximization algorithm and transform the data into approximately stationary
series with the outliers replaced, so that we obtain a total of $T=785$
monthly observations for each macroeconomic variable. One can refer
to \cite{McCracken2016} for the detailed data description.

In selecting the number of breaks, we set the minimum segment length to $h=20$ months and restrict candidate break dates to the monthly grid. We consider $m\in\{0,1,\ldots,8\}$ as the candidate numbers of breaks. For each $m$, we minimize the QML objective over all admissible partitions using the dynamic programming algorithm and then select $\hat m$ by minimizing
Eq.~(\ref{model_selection}).
In our empirical application, $N=126$, $r=7$ (selected by Bai--Ng IC2 on the full sample), and $\hat\rho=0.9574$, so the penalty term is numerically
\[
\log(\min\{N,T\})\,r^2(1+|\hat\rho|)=\log(126)\times 7^2\times (1+0.9574)\approx 463.86.
\]
Hence the penalty used in the information criterion is $463.86\times m$. No additional tuning constant is used. The selected model yields five breakpoints: January 1969, January 1983, June 2008, March 2010, and February 2020. These breakpoints correspond to significant economic
events. January 1969 marked a surge in U.S. inflation, primarily driven
by increased spending on the Vietnam War and the ``Great Society''
programs. January 1983 roughly coincides with the onset of the Great
Moderation, a period characterized by reduced macroeconomic volatility.
June 2008 signifies the early stages of the Global Financial Crisis,
as the collapse of the housing market began to impact the broader
financial sector. March 2010 is approximately the end of the Great
Recession for the U.S. economy. February 2020 represents the initial
economic disruptions stemming from the COVID-19 pandemic, which rapidly
led to a sharp global downturn.

Using the estimated breakpoints, we split the sample into six regimes and apply Bai--Ng IC2 within each regime, obtaining 2, 5, 7, 3, 4, and 7 factors, respectively.
Under our classification rule, the variation in the estimated numbers of factors across adjacent regimes suggests that all identified breaks are Type 1 changes.
For clarity, the regime with three factors is 2008:07--2010:03, whereas the following regime 2010:04--2020:02 contains four factors. Since principal components are identified only up to rotation, the economic meaning of individual factors is not always clear, and higher-order components are particularly difficult to interpret. We therefore focus on a more stable feature of the data: the concentration of explained variance in the leading principal components.

This concentration rises substantially in the crisis regimes. In 2008:07--2010:03, the first principal component explains about $34.7\%$ of total variance and the first three explain about $59.4\%$. In 2020:03--2024:07, the first principal component alone explains about $52.6\%$ of total variance. By contrast, in the non-crisis regimes 1983:02--2008:06 and 2010:04--2020:02, the first principal component explains only about $10.0\%$ and $13.2\%$, respectively. This indicates that crisis periods are dominated by a small number of strong common shocks, whereas common variation is more diffuse in normal times.

The two crisis episodes also differ in the composition of the leading common variation. During 2008:07--2010:03, the leading common variation is more closely associated with financial and price-related categories, including money and credit, interest rates and exchange rates, and prices. During 2020:03--2024:07, the leading common variation is much more strongly concentrated in output and income, as well as labor-market variables. This difference is consistent with the distinct nature of the Global Financial Crisis and the COVID-19 shock. Detailed regime-wise explained-variance results are reported in Appendix E of the supplementary material.

\vspace{-1em}
\section{Conclusions}
\vspace{-1em}

This paper introduces a quasi-maximum likelihood
(QML) method for estimating multiple breakpoints in high-dimensional
factor models. We analyze two distinct types of structural changes
and develop corresponding asymptotic theories for the QML estimators.
For Type 1 changes, we establish the consistency of the QML estimator,
demonstrating that it converges to the true breakpoint with probability
approaching one as the sample size increases. For Type 2 changes, we
show that the distance between the QML estimators and the true breakpoints
is stochastically bounded (fractionally consistent).

The proposed QML method is straightforward
to implement and computationally efficient, requiring only a single
full-sample eigendecomposition. Furthermore, we demonstrate that the
number of breaks can be consistently estimated using our information
criterion. Extensive simulation results confirm the strong finite-sample
performance of our approach. Finally, an application to the FRED-MD
dataset underscores the method's practical utility in empirical macroeconomic
analysis.

\vspace{-1em}
\section*{Supplementary materials}
\vspace{-1em}

The proofs for the theorems and propositions are given in Appendix A of the supplementary material, the proofs for the technical lemmas are given in Appendix B, sufficient conditions for Assumptions \ref{error_sup} and \ref{Central_Limit} are provided in Appendix C, additional simulation results are provided in Appendix D, and additional empirical results are reported in Appendix E.

\vspace{-1em}
\section*{Data availability statement }
\vspace{-1em}

The data that support the findings of this study are openly available at:\\
\url{https://www.stlouisfed.org/research/economists/mccracken/fred-databases}.

\renewcommand{\bibfont}{\small}
\spacingset{0.8}
 \bibliography{bibliography}

\end{document}